\documentclass[twocolumn]{aastex701}
\usepackage[utf8]{inputenc}
\usepackage{graphicx}
\graphicspath{{Images/}}
\usepackage{float}
\usepackage{xcolor}
\usepackage{amsmath}

\usepackage{natbib}

\newcommand{\degree}{$^\circ$}
\usepackage{enumitem}
\newcommand{\SNR}{\ensuremath{\text{SNR}}}
\newcommand\thesis[1]{}

\newcommand{\DATAVOL}{371~GB }

\newcommand{\SHanallyzed}{\ensuremath{h_{ps}=1128~\text{source-hours} }}

\newcommand{\Nstarsanalyzed}{15,029}

\newcommand{\NWDanalyzed}{217}
\newcommand{\Nquasarsanalyzed}{10}

\def\ie{{\it i.e.,}}
\def\eg{{\it e.g.,}}
\let\longtable\relax
\let\endlongtable\relax

\begin{document}


\title{Probing Millisecond Optical Timescales with Continuous-Readout Mode Observations from ZTF}

\author[0009-0006-4823-9768]{Shar Daniels}\affiliation{Department of Physics and Astronomy, University of Delaware, Newark, DE 19716, USA}\email{shard@udel.edu}
\author[0000-0003-1953-8727]{Federica Bianco}\affiliation{Department of Physics and Astronomy, University of Delaware, Newark, DE 19716, USA}\affiliation{Joseph R. Biden, Jr. School of Public Policy and Administration, University of Delaware, Newark, DE 19716, USA}\affiliation{Data Science Institute, University of Delaware, Newark, DE 19716, USA}\affiliation{Vera C. Rubin Observatory, Tucson, AZ, USA}\email{fbianco@udel.edu}
\author[0000-0002-8977-1498]{Igor Andreoni}
\affiliation{Department of Physics and Astronomy, University of North Carolina, Chapel Hill, NC 27599, USA}\email{igor.andreoni@gmail.com}
\author[0000-0003-2242-0244]{Ashish Mahabal}\affiliation{Division of Physics, Mathematics and Astronomy, California Institute of Technology, Pasadena, CA 91125, USA}\affiliation{Center for Data Driven Discovery, California Institute of Technology, Pasadena, CA 91125, USA}\email{aam@astro.caltech.edu}
\author[0000-0002-5884-7867]{Rich Dekany}\affiliation{Caltech Optical Observatories, California Institute of Technology, Pasadena, CA 91125, USA}\email{rgd@astro.caltech.edu}
\author[0000-0002-9276-3261]{Gregory Dobler}\affiliation{Department of Physics and Astronomy, University of Delaware, Newark, DE 19716, USA}\affiliation{Joseph R. Biden, Jr. School of Public Policy and Administration, University of Delaware, Newark, DE 19716, USA}\affiliation{Data Science Institute, University of Delaware, Newark, DE 19716, USA}\email{gdobler@udel.edu}
\author[0000-0002-3168-0139]{Matthew Graham}\affiliation{Cahill Center for Astrophysics, California Institute of Technology, Pasadena, CA 91125, USA}\email{mjg@caltech.edu}
\author[0000-0001-7062-9726]{Roger Smith}\affiliation{Caltech Optical Observatories, California Institute of Technology, Pasadena, CA 91125, USA}\email{rsmith@astro.caltech.edu}
\author[0009-0005-3569-9944]{Rujula Yete}\affiliation{Department of Physics and Astronomy, University of North Carolina, Chapel Hill, NC 27599, USA}\affiliation{Department of Astronomy, University of Maryland, College Park, MD 20742, USA}\email{rjyete@umd.edu}


\begin{abstract}

The subminute-timescale phase space is an underexplored regime of the optical sky. Sub-second optical observations are difficult to accomplish with traditional CCD exposure-readout cycles, and alternative technologies like EMCCDs and CMOS have historically limited surveys to targeted observation. 
Continuous-readout mode enables millisecond-timescale observations on traditional cameras by leaving the shutter open during readout, enabling high-cadence observations on all-sky surveys. We present the first results from continuous-readout mode data from the Zwicky Transient Facility. We developed a deep-learning-based data analysis pipeline and report results on 141 square degrees, or \SHanallyzed, of 
data, sampled at 300x/second. Our Convolutional Neural Network can retrieve optical transients of duration 9.7-29.2 milliseconds with a precision of $66\%$ and recall of $78\%$ for brightness changes of 10\% and greater, and our subsequent light curve-based analysis has an efficiency of $>94.6\%$. 
We retrieve a total of $\sim300,000$ candidates on sources of magnitude $G_{RP}< 12$, of which 100 are above $\SNR{}=1.55$ and were visually inspected.
After accounting for noise and filtering contaminants both known and unique to this mode of observation, these 100 candidates were eliminated as being astrophysical in nature, leading to
upper limits on the observed rate of $\sim10$~ms transient events associated with
point sources as well as specifically with white dwarfs and quasars. This first survey-scale characterization of the bright millisecond optical transient and variable sky establishes the rarity of short-timescale variability in the optical regime.

\end{abstract}


\section{Introduction}

Transient astronomy reveals extreme astrophysical phenomena dominated by fundamental and high-energy physics. Astrophysical transients can 
be described by the timescale over which they 
vary and by their energy output. The number of discovered time-evolving phenomena has exploded since the beginning of the 20th century. 

\begin{figure}[h!]
  \centering
  \includegraphics[width=1\linewidth]{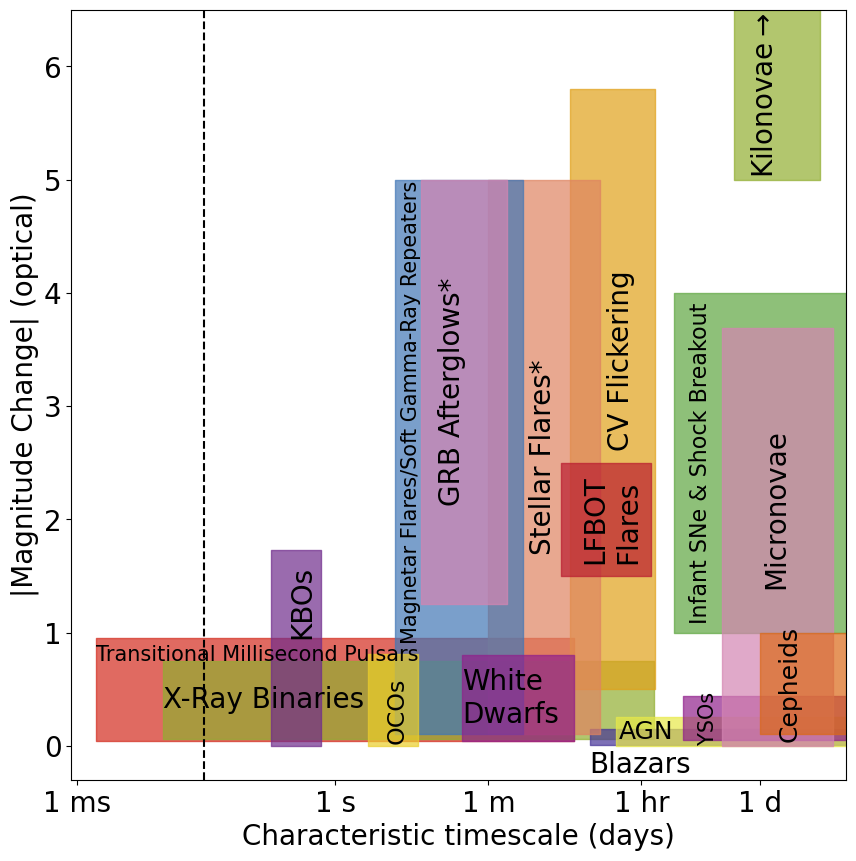}
  \caption{Optical fast transient phenomena categorized by characteristic timescale and absolute magnitude change. An asterisk denotes that the shown time scale is for the rise time of the phenomena, as it is the most scientifically valuable to capture observations during this period. The discovery space is apparent below the time scale of $\sim1$ day. This region of parameter space is difficult to explore in optical wavelengths because of the typical duration of an expose-readout CCD cycle. KBOs: Kuiper Belt Objects; OCOs: Oort Cloud Objects; GRB: Gamma Ray Burst; LFBOT: Luminous Fast Blue Optical Transient; CV: Cataclysmic Variable; AGN: Active Galactic Nuclei; SNe: Supernovae; YSOs: Young Stellar Objects. For more information on all these sources, see \autoref{sec:phasespace}, \autoref{tab:phasespace}. This work searches the left-hand side of this plot, up to the dotted line drawn at $\sim30$ ms.}
  \label{fig:phasespace}
\end{figure}

While over the past few decades we have discovered a wealth of new phenomena evolving at time scales from years to hours, the optical sky's ``transient zoo'' is minimally populated at minute-to-sub-second, in part 
because traditional optical observational methods cannot 
access these timescales (see \autoref{fig:phasespace} for fast optical transient phenomena that are currently known).

Several phenomena
are expected to generate observable brightness changes on second and sub-second timescales: stars
occulted by solar system objects \citep{nihei2007detectability}, episodic bursts of cataclysmic variable stars \citep{bruch2021comparative}, activities in
blazars \citep{raiteri2021dual}, and potentially even counterparts to Fast
Radio Bursts (FRBs) \cite[\eg{}][]{chen2020multiwavelength}. Importantly, this underexplored region of the parameter space may also contain
entirely new and unexpected classes of objects. However, many phenomena at these time scales are also typically rare. For example, Kuiper Belt occultations, whose typical time scales are $~0.1$ seconds \citep{nihei2007detectability}, are estimated at $10^{-9}$ occultations in any given moment \citep{nir2023search}. Over the past decade, a contaminant foreground phenomenon happening on these time scales has emerged and grown: glints from satellite constellations and space junk \citep{2016RMxAC..48...91K, 2021MNRAS.505.2477N, 2023CoSka..53d..69K, 2020ApJ...903L..27C, 2025ApJ...994..175T}. The rapid growth of satellite constellations \citep{doi:10.1126/science.adi4639}, and therefore of the associated glints, makes searches in this region of the parameter space all-the-more timely.

Readout noise would typically dominate observations with traditional CCDs at subsecond timescales, and the open-close shutter cycle is itself typically longer than a few seconds. Instrumentation built specifically for astrophysical observations on sub-second timescales exists, including frame transfer CCDs and 
CMOS (Complementary Metal-Oxide-Semiconductor), which enable rapid readout with low read noise. Until recently these devises were, however, prohibitively expensive compared to traditional CCDs, historically limiting applications to small field of view surveys and, therefore, targeted observations for rapid transient detection \citep[\eg{} ][]{dhillon2007ultracam, 2008AJ....135.1039B} and lucky imaging \citep[\eg{}][]{2006A&A...446..739L, 2012SASS...31..147G}. Modern CMOS offer a low-cost solution for high-speed photometric surveys and several projects using them are emerging. The Organized Autotelescopes for Serendipitous Event Survey observed $\sim$2,000 stars with magnitude V $\lesssim$ 13.0 to search for transient sources with durations of 0.2-2 s, finding none so far \citep{arimatsu2017organized, arimatsu2021detectability}. The TAOS II survey is collecting 20 Hz CMOS observations for 2 continuous hours per field, targeting ecliptic fields to detect Kuiper Bet and Oort Cloud objects \citep{huang2021taos}. 
The Tomo-e Gozen instrument on the KisoSchmidt telescope is undertaking the Northern Sky Transient Survey, observing 12,000 square degrees per night with CMOS readout times of 2 frames per second \citep{sako2018tomo, zhang2024optical}. The Weizmann Fast Astronomical Survey Telescope can observe with CMOS at a frame rate of up to 90 Hz, while nominal observations are conducted at $10–25~\mathrm{Hz}$ \citep{nir2021weizmann}.
 The Colibri photometry array will use electron-multiplied charge-coupled devices (EMCCD) to obtain a sampling rate of 40 Hz \citep{pass2017pipeline}. 
 Finally, Everyscope \citep{law2014evryscope} and its successor, the upcoming Argus Array \citep{law2022low}, are wide sky deep and rapid surveys with arrays of small telescopes, and are poised to revolutionize the field. These are all specialized surveys, purposefully designed for high speed photometry.

Modified observing modes with traditional cameras, like trailing \citep{howell1986time, tingay2021high} and continuous-readout \citep{bianco2009search}, offer an alternative to specialized instrumentation that allows observations as fast as millisecond timescales on most cameras. This includes large field-of-view surveys, which can lead to vast amounts of collected data. Because transient phenomena are often rare, the ability to scour large areas simultaneously with a large field of view surveys is critical to discovery. In trailing mode, Tingay 2021 found an upper limit to the rate of 21-ms transients of 0.8 per square degree per day at V magnitude $\leq6.6$ \citep{tingay2021high}.



\begin{figure}[h!]
  \centering
  \includegraphics[width=1\linewidth]{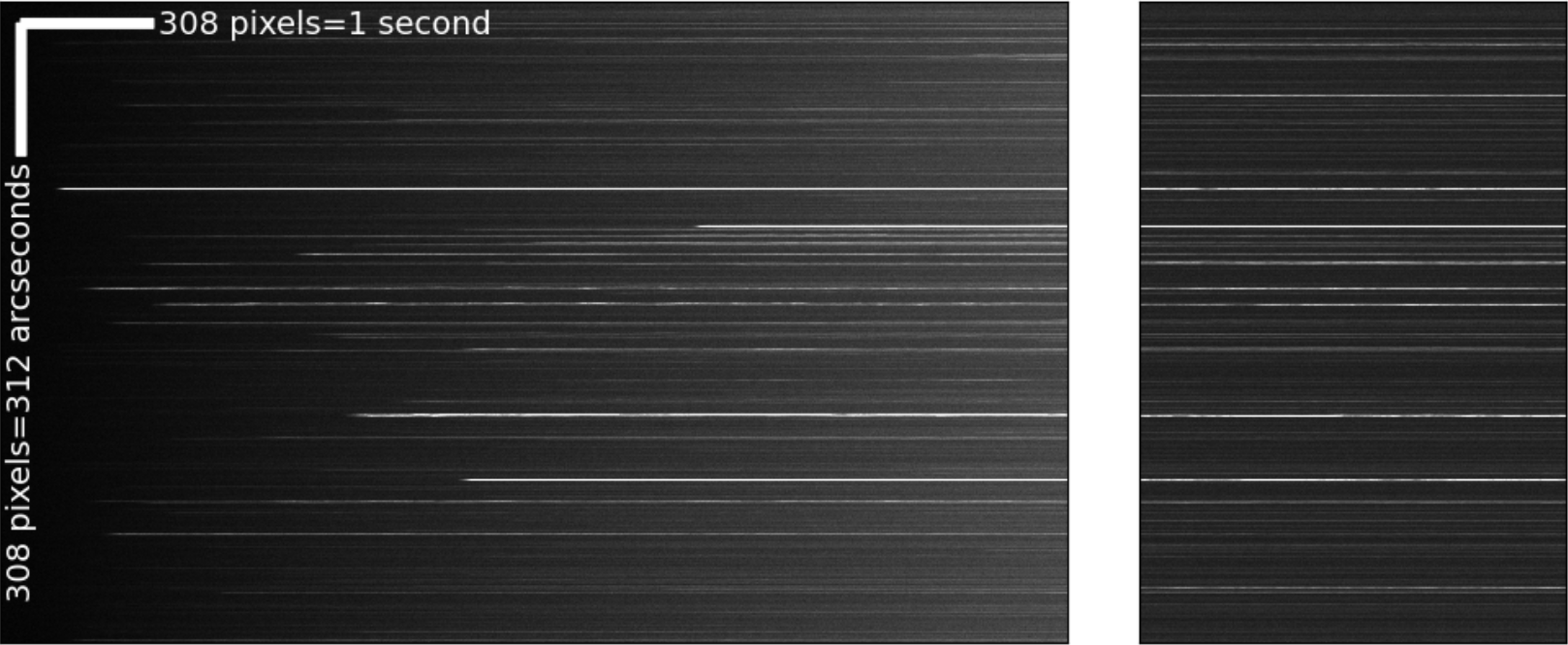}
  \caption{Continuous-readout mode imaging. Left: the beginning of the observing run and image file. The streak starts at the CCD position of the source (star/quasar/white dwarf) in a standard read-out image along the $x$ (time) direction in the leftmost panel. Right: a segment taken from the middle of the same continuous-readout run where all sources have developed into streaks. The spatial and time pixel scales are indicated in the graphic. To enhance visibility, the brightness scale is set to the log-normal of the pixel counts and clipped at 0.15 and 0.25 for the left and right panels, respectively.}
  \label{fig:combined}
\end{figure}

Trailing and continuous-readout modes enable transient discovery at sub-second timescales on traditional cameras by leaving the shutter open during slew or readout respectively \citep{bianco2009search}. This integrates each astrophysical image along one spatial dimension so that the resulting image has one time dimension (the horizontal direction in \autoref{fig:combined}) and one spatial dimension (here, the vertical direction). Stars and quasi-stellar objects are smeared into time-resolved streaks along the time dimension. 
These methods allow us to resolve the data at sub-second rates without 
incurring additional costs of specialized equipment, and make it possible for us to take advantage of synoptic 
survey telescopes with large fields of view such as the Zwicky Transient Facility \cite[ZTF;][]{bellm2019, Graham2019}.

Continuous-exposure imaging has been used in the past to set limits on the size distribution of Kuiper
Belt objects \citep{bianco2009search, bianco2010taos, zhang2013taos}. We have carried out the first analysis of continuous-exposure images with deep
learning. The objective is to build an efficient method to detect transients at scale with both expected and
unexpected signatures in a continuous-exposure survey. Neural networks are excellent
analysis tools for this purpose because they can handle large amounts of data and find subtle and rare patterns within.

The paper is organized as follows: We present the continuous-readout mode ZTF survey in \autoref{sec:data}, our convolutional neural network (CNN) training and results in \autoref{sec:cnn_everything}, our post-CNN processing in \autoref{sec:postcnn}, and our resulting transient candidates in \autoref{sec:candidates}.

\section{The Continuous-Readout Mode ZTF Survey}\label{sec:data}

\begin{figure}[h!]
  \centering
  \includegraphics[width=\columnwidth]{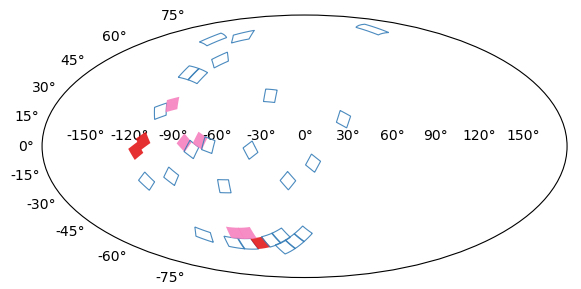}
  \caption{Galactic coordinate map showing the location of the fields observed with ZTF in continuous-readout mode. Each field covers 47 square degrees, the field of view of the 600-megapixel ZTF CCD mosaic. The model was developed and trained on data from the fields marked with pink and red; the results reported in this work refer to the fields marked with red. Blue outlines are additional fields where the survey has observed in continuous-readout mode to date. }
  \label{fig:crm}
\end{figure}

We conducted a pilot survey in continuous-readout mode at the Palomar Telescope Samuel Oschin robotic telescope (P48) as a special program using the ZTF camera. A total of \DATAVOL, or \SHanallyzed{} (one hour of observations of one point source), were collected between 2022-08-20 and 2022-10-26 and analyzed as part of this work. In the ZTF continuous-readout mode, the observations are organized in sequences stored in FITS files, each typically 3,080 pixels along the time axis.  The sampling rate (\ie{} time of charge-scrolling by one pixel) is approximately 300 Hz. However, adjacent observations are not independent due to the finite PSF of the sources, which is typically of the order of $\text{FWHM}\sim2$~pixels = 6 ms. These data cover 14 fields, both extragalactic and Galactic, as described in \autoref{tab:fields} and \autoref{fig:crm}. Each field is 47 square degrees. Since for all of the observations analyzed here the sun elevation was $\mathrm{Sun}_{el.}\lesssim$ -28\degree, we expect contamination from satellite glints to be minimal.

In this pilot study, we analyzed data from 248 ZTF full-frame images. They were organized in 3842 FITS files, one per CCD, each one with four extensions - an extension for each quadrant of the CCD. We analyzed each one of the 15,368 quadrants individually as a separate image. All results reported here are derived from the analysis of data collected on 2022-10-20, and correspond to the red fields shown in \autoref{fig:crm} and bolded in \autoref{tab:fields}. With exploratory cuts on the brightness of the streaks, this work includes the analysis of \SHanallyzed{} from approximately \Nstarsanalyzed{}  stars, \Nquasarsanalyzed{} quasars, and \NWDanalyzed{} white dwarfs.\footnote{Only the approximately ten brightest sources for each time-resolved image were selected for this pilot work, each of magnitude $G_{RP}< 12$ (see \autoref{sec:crossmatching}).}

\begin{table*}[t]
\centering
\footnotesize
\begin{tabular}{|c|c|c|c|c|c|c|c|c|c|c|}
\hline
\textbf{Field } & \textbf{RA}& \textbf{Dec} & \textbf{Gal. Long}& \textbf{Gal. Lat}& \textbf{Ecl. Lat}&\textbf{Duration } & \textbf{Images } & \textbf{Filter}& \textbf{Date} & \textbf{Sun el.}\\ 
\textbf{ID} & (deg) & (deg) & (deg) & (deg) & (deg) & (s) & (\#) &  & (UT) & (deg)\\ \hline

 798& 279.99351 & 54.95 & 84.3050 & 23.3794 & 77.44& 60 & 384 & $r$& 2022-08-20 05:41:14 & -34.76\\ \hline
 
802 & 324.38344 & 54.95 & 97.7770 & 1.9883 & 62.14& 308 & 1976 & $i$ & 2022-10-14 04:03:38& -35.54\\ 
 
 & &&& & & 308 & 1976 & $g$ & 2022-10-14 03:54:35& -33.71\\  \hline

501 & 15.74212 & 11.75 & 128.4030&-50.7038 &4.66&
308& 1972 & $g$ & 2022-10-14 06:32:33 & -61.13\\
 
 & &&& & &308 & 1972& $r$ & 2022-10-14 06:41:00& -62.04\\ \hline

831 & 340.0 & 62.15 & 108.5417&3.2177&61.02&
620 &3968& $i$  & 2022-10-17 03:39:29& -31.38\\ 
 
 & &&& & & 308 & 1976 & $g$ & 2022-10-17 03:48:37& -33.26\\ \hline
 
500 & 8.75787 &11.75 &117.5970&-50.7038& 7.33&
930 &5952& $r$  & 2022-10-17 06:04:30& -58.47\\ \hline

 



\textbf{686 } & 298.26936 &33.35 & 69.5025&2.7706&52.85&
620&3968& $i$  & 2022-10-20 03:19:05& -27.87\\ 
 
 & &&& & &310 &1984 & $g$ & 2022-10-20 03:33:11& -30.80\\ \hline

\textbf{640 } & 300.7568 & 26.15 & 64.5379&-2.8742&45.33&
620&3968& $i$ & 2022-10-20 03:41:57& -32.61\\ 
 
 & &&& & &310 &1984 & $g$ & 2022-10-20 03:55:03& -35.30\\ \hline

\textbf{449 } & 19.26088 & 4.55 & 136.0305&-57.3406& -3.33&
620&3968& $r$  & 2022-10-20 07:03:46& -66.09\\ \hline

\end{tabular}
\caption{Table describing field pointings used in this work. Results are obtained from the analysis of the fields shown in bold. In addition, our CNN was trained on data from all fields observed on the nights of 2022-10-17, 2022-10-14, 2022-08-20, and 2022-10-20. Each image is taken over 10 seconds. Sun elevation (Sun el.) is included as it directly impacts the probability of contamination from satellite glints.} 
\label{tab:fields}
\end{table*}

\section{CNN-based Continuous-Readout Image Analysis}\label{sec:cnn_everything}
\subsection{CNN Architecture}\label{sec:cnn}

\begin{figure}[h!]
\centering
\includegraphics[width=1\linewidth]{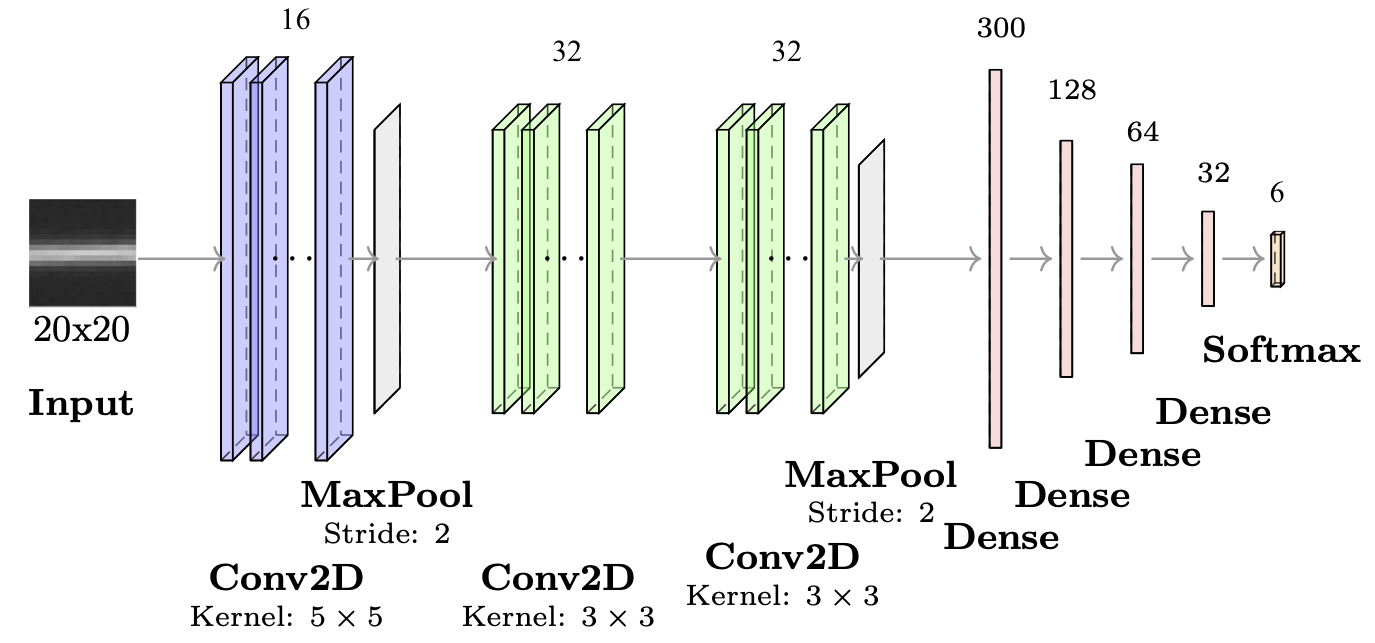} 
\caption{Architecture of the CNN built to analyze ZTF continuous-readout data. The model has four convolutional layers followed by a Multilayer Perceptron (MLP) classification head. The model hyperparameters are reported in \autoref{tab:hyperparams}. We used the Adam optimizer and categorical cross-entropy loss.}
\label{fig:cnn}
\end{figure}

Since the morphology of our sources is distinct from a classical 2D PSF source or extended object profile, custom-built analysis tools are necessary to analyze continuous-readout mode data. Neural networks are excellent analysis tools for this purpose because they can handle large amounts of data and find subtle and rare patterns within\thesis{. Inspired by biological neural systems, neural networks are supervised machine learning models. Given examples of input-target pairs, they are ``trained'' to associate input features with target labels. Neural networks consist of sequential layers of interconnected nodes, called ``neurons'', with trainable weights. In the most generic neural networks, Multilayer Perceptrons, the neuron performs a multilinear regression on the input given by the previous layer, the result of which is then passed to a non-linear operator called an ``activation function'', enabling the model to learn complex non-linear patterns} \citep{gurney2018introduction}. Convolutional Neural Networks (CNNs) can learn, via learnable convolutional kernels as neurons, patterns in adjacent input features, thus excelling in recognizing patterns in image-like data where the features are pixel values \citep{li2021survey}.
We designed and deployed a patch-based classification Convolutional Neural Network (CNN). That is: each pixel in each analyzed image is classified in the context of a window of $20\times20$ pixels around it (a ``postage-stamp''). The CNN returns a probability that the pixel at the center of the postage stamp belongs to each of the $N_c$ classes. The architecture of our CNN is shown in \autoref{fig:cnn} and \autoref{tab:hyperparams}. 

\begin{figure*}[t!]

\centering
\includegraphics[width=.8\linewidth]{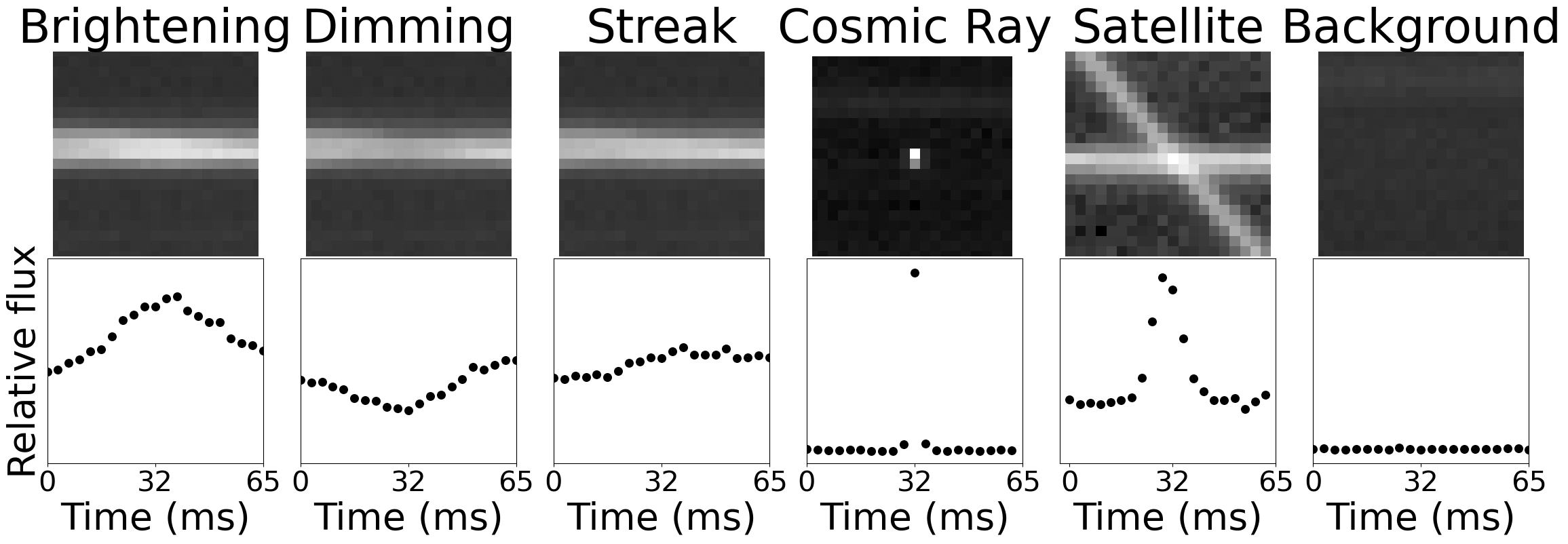} 

\caption{An example of each of the six classes that the CNN was trained to categorize. The top panels show the postage stamp of image data, the input to the CNN; the bottom row shows the corresponding light curve, extracted with aperture photometry (as described in section \autoref{sec:lightcurveextraction}). 
Brightening and dimming transients were implanted. Cosmic rays and satellites were a source of contamination in early detection trials and were selected by visual inspection of early CNN outputs. 
The CNN was also trained on 
streaks and background examples. The brightening stamp shown has amplitude $A=1.3$ and duration $w=5$ pixels (16 ms); the dimming stamp $A=.7$ and $w=4$. See \autoref{sec:injections} and \autoref{eq:gaussian}.}
  \label{fig:classes}
\end{figure*}

\begin{figure}[h!]
  \centering
  \includegraphics[width=1\linewidth]{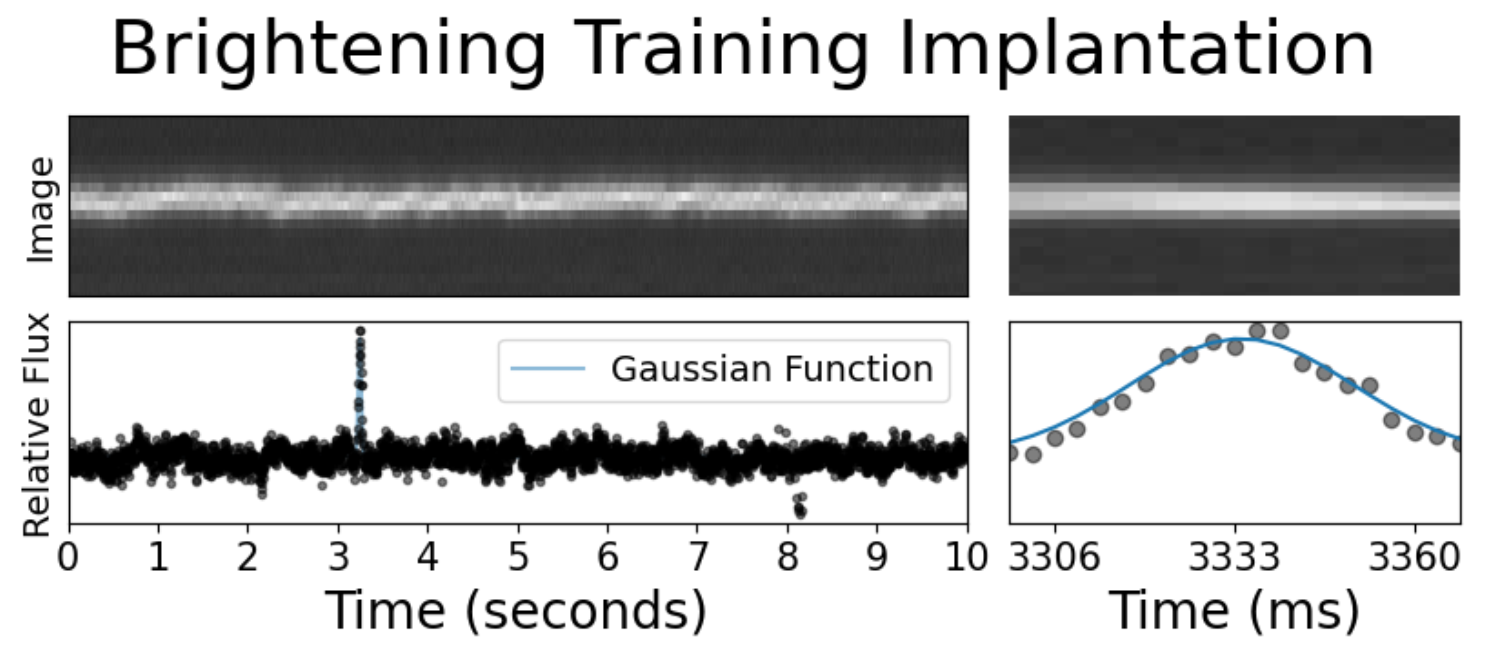}
  \caption{Top: a streak with an implanted transient. Bottom: the resulting light curve. The right panels show a zoomed-in region of the same data, centered on the implanted event. This implanted transient has amplitude $A=1.3$ and $w=16 $ ms (5 pixels). The event is centered at $\mu=3,333$ ms (pixel column 1,000). See \autoref{sec:injections} and \autoref{eq:gaussian}.}.

  \label{fig:gaussian_implant}
\end{figure}
The objective of our analysis is to retrieve transient events that can manifest as both brightening and dimming of the streak on millisecond time scales.
However, to ensure our model can accurately recognize the events of interest, we also had to train it to recognize all patterns in the data, including basic data structures and potential contaminants. Thus, we created examples of each class of interest and trained our CNN to distinguish $N_c=6$ classes: streaks, streaks with a brightening transient, streaks with a dimming transient, cosmic rays, satellites, and background. \thesis{Thus the output of our model is a $6D$ probability vector $\vec{P}$ for each pixel in the image, containing a normalized score for each class, which can be interpreted as a probability of each class.} 


\begin{figure*}[t!]
  \centering
  \includegraphics[width=0.8\linewidth]{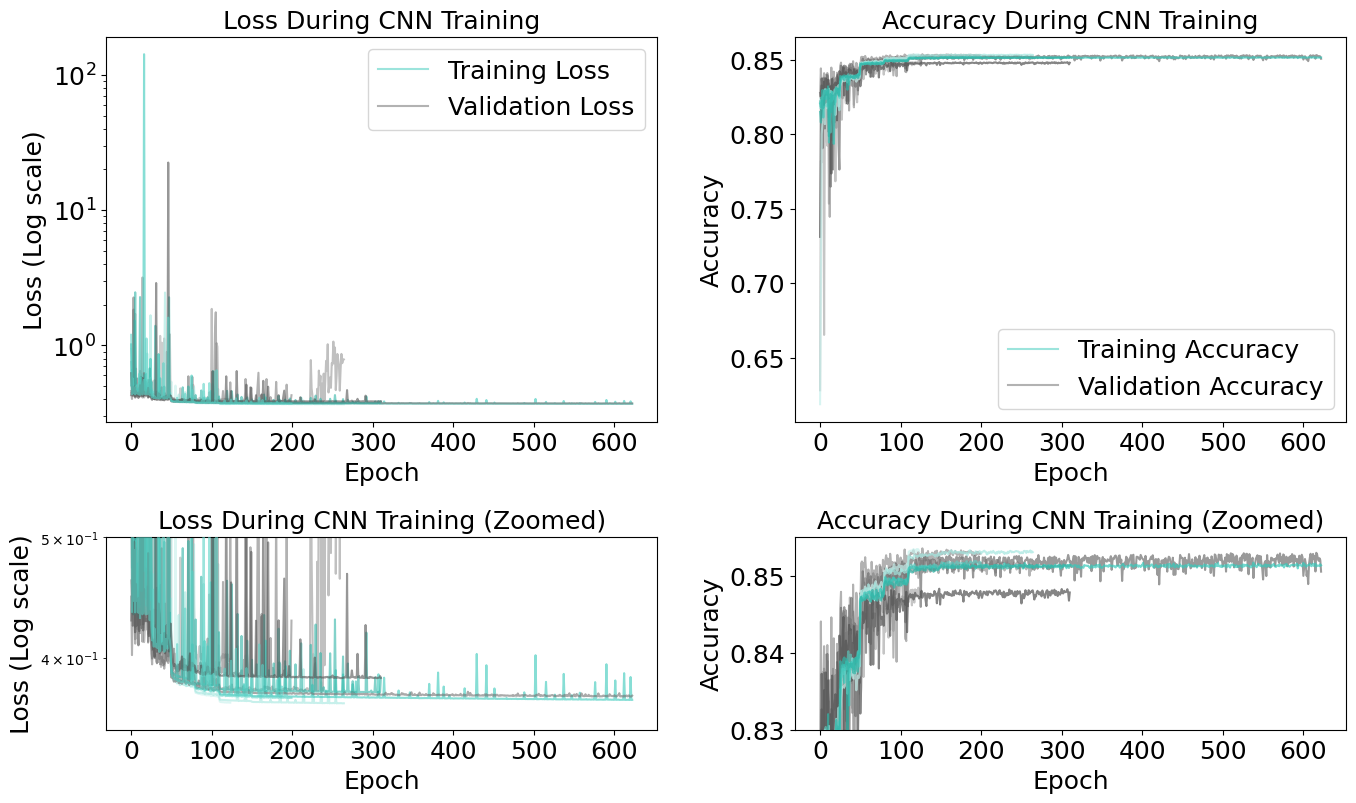}
  \caption{CNN training history. The loss (categorical cross-entropy) and accuracy are reported for both training and validation sets. The learning rate was modified during training at $\mathrm{epoch} = 25, 50, 80, 110,$ and $150$, to reduce the noise in the validation loss's impairment of learning, which produces the plateaus in the loss function. Each fold is plotted as a different shade of teal for training and a corresponding shade of grey for validation.}
  \label{fig:lossacc}
\end{figure*}

\subsection{Image Preprocessing and Construction of a Training Dataset}\label{sec:injections}

The images are pre-processed to make the data suitable for applying the CNN. First, we subtract the background from each image, where an ``image'' represents one continuous observing run from a single CCD quadrant. The background is calculated as the mean of the image in an iterative 3-sigma clipping loop that converges when the mean is stable to 1e-3.
\thesis{that is: as a constant for each image by iteratively clipping the maximum of each image array at a 3 standard deviation level until there is a $<0.0001$ change between iterations. }
Following background subtraction, images are min-max scaled\thesis{, meaning that per each image file the pixel with the highest value is set to 1 and the lowest to 0. This is a common choice for preprocessing image data for CNN-based analysis as it is observed empirically that CNNs perform better when the input data is near unity}. Then, a $20\times20$ pixel postage stamp centered on each pixel is saved as a separate array. For training, some stamps are then implanted with transients, as described below.


To train the CNN, postage stamps corresponding to streaks, cosmic rays, background, and satellite segments were pulled directly from the data. The first five categories of stamps were sourced from 1,148 FITS files, and the satellites were sourced from eight additional FITS files.

{\it Streaks stamps:} One central stamp was taken from each of the 10 brightest streaks per image (quadrant of the FITS file). The placement of the stamps along the streak in the time axis (\ie{} row) was chosen following a uniform random distribution with the constraint that the edge of each stamp was not within 10 pixels of the edge of the image. Four additional streak stamps were taken next to each central stamp: two on each side, to cover the full width of the streak and to ensure that the CNN could categorize pixels on the edges of streaks. In total, the training and testing set included 57,400 streak stamps.

{\it Background stamps:} Anything $>$3 pixels away from the center of a streak was considered to be a background stamp, and background stamps were chosen at a distance of 4-25 pixels in the spatial direction from the center of every streak, brightening transient, and dimming transient. We 
randomly selected
114,800 such postage stamps for training and testing.

{\it Cosmic rays:} Cosmic rays were found to be a significant source of contaminants in testing early versions of our CNN; 600 postage stamps containing cosmic rays were saved to be included in the training data. 

{\it Satellite stamps:} Satellites were found to be a source of contaminants in testing early versions of our CNN; 72 intersections of satellites and streaks were retrieved, from 10 distinct satellites. For each of these 72 intersections, every pixel $<$ 10 pixels away from the center of the intersection either row-wise or column-wise was saved as the center of a ``satellite" postage stamp, leading to 25,460 satellite stamps.

{\it Transient stamps:} Brightening and dimming transient stamps were created by modifying the streak stamps, leading to 57,400 stamps with brightening transients and the same number with dimming transients. The transients were injected by modifying the pixel brightness by a Gaussian multiplicative factor for pixels within the existing streak. The boundaries of streaks 
were identified by the following criteria: the brightest average row (where rows are along the time axis) in the postage stamp and the rows on either side of it are included, and rows that are 2-6 pixels away from the brightest are checked sequentially and included if they are both brighter than 10\% of the brightest row and dimmer than the adjacent row closer to the center of the streak (to avoid counting two overlapping streaks as one streak). For each selected row along the time axis, we then multiply the pixel value $F$ by a Gaussian as described in \autoref{eq:gaussian}:

\begin{equation}
  F_{event} = F \times \left((A-1)e^{ -\frac{(y-\mu)^2}{2w^2} }+1\right)
  \label{eq:gaussian},
\end{equation}


\noindent where $A$ is the amplitude, $\mu$ is the event center pixel on the time axis, $y$ is the pixel number corresponding to the time axis, $w$ is the standard deviation of the Gaussian model, or duration of the event.\footnote{Note that we added 1 since this is a multiplicative injection. 
That is: $ \lim_{(y - \mu)^2 \to \infty} F_{event} = F \times 1$.}

 The duration of the brightening and dimming transients is characterized by the standard deviation $w$ of the multiplicative Gaussian factor and set in the range [3-9] pixels ($\sim10-30$ milliseconds) with the same number of events at each duration. The brightening and dimming factors were tuned to improve performance for transients of lower intensity by sampling the amplitude $A$ range more finely as values approach $A=1$. Within the range $A$=[1.005, 2] for brightening events ($A$ = [0.5, 0.995] for dimming) for every amplitude $A$ as reported in \autoref{tab:bright}, 
 1,275 brightening transients (1,435 dimming transients) were implanted at each amplitude.

\begin{table}[h!]
\centering
\begin{tabular}{c|c}

\multicolumn{2}{c}{\textbf{Implantation Amplitude $A$}}\\
\hline
\textbf{ Brightening } & \textbf{Dimming } \\

\hline\hline

1.005 & .995  \\ 
1.01 &  .99 \\ 
1.05  & .95 \\
1.15  & .9 \\ 
1.2  & .8 \\ 
1.3  & .7\\
1.4  & .6\\ 
1.5  & .5\\ 
2  & \\ 

\end{tabular}
\caption{Amplitudes ($A$), the multiplicative factors in \autoref{eq:gaussian}, used for synthetic brightening and dimming event implantations in the training and test sets. }
\label{tab:bright}
\end{table}

An example of a stamp from each of the five training classes is shown in \autoref{fig:classes}, and an example of the Gaussian implantation in the whole streak is shown in \autoref{fig:gaussian_implant}.

The process described above produced a significant class imbalance with 57,400 examples of steady streaks, brightening events, and dimming events, 114,800 examples of background, 25,460 satellites, and 600 cosmic rays. Class imbalance is always a source of concern in classification as models can learn that predictions associated with the most common class(es) are less risky, resulting in a bias. 
However, as we will show in \autoref{sec:metrics}, our least numerous class, namely cosmic rays, had high recall. 
Conversely, the background may present with a variety of brightness patterns, justifying the larger training sample size.

While, as we argued in the introduction, the optical universe at these time scales is underexplored, a few phenomena are expected or predicted at similar amplitude and time scales of those we implanted. Particularly, the event depicted in \autoref{fig:gaussian_implant} is consistent with observed X-ray binary pulsator events (which however would be repeated) \citep[\eg{}][]{2018ATel11426....1S} or a satellite glint, the reflection of sunlight from small satellite components, which would be serendipitously aligned with the source. Following the parametrization in \citep{nir2021weizmann}, which assumes a satellite is on a circular orbit observed at an elevation angle of 47.9\degree\ which gives a distance $R_\mathrm{sat} = 37,200$~km and an albedo of 4\%, with the magnitude of the stars where the event was implanted at M=12, the size of the satellite component producing the glint would be 7~cm. Additional, as of yet unobserved but theoretically predicted phenomena include
the very early prompt optical counterpart (POC) of a GRB \citep[][although we implanted this event on a star, while we would expect these phenomena to be not associated with a streak or to be associated with the streak produced by a galaxy]{panaitescu2022properties}; an asteroid size primordial black hole (PBH) could produce a similar event \citep[][although the transit velocity of the PBH across the background source would need to be significantly higher than the rotational velocity of an object in the Galaxy]{montero2019revisiting}.

\begin{table}[h!]
\footnotesize
\centering
\begin{tabular}{c|c|c|c|c|c}

\textbf{Layer} & \textbf{Kernel } 
& \textbf{Param}& 
\textbf{Pad-} & \textbf{Acti-}& \textbf{Regu-} \\
\textbf{Type} & \textbf{Size} 
& \textbf{(\#) }& 
\textbf{ding} & \textbf{vation}& \textbf{larizer} \\ 

\hline
Conv2D & (5,5) 
& 416 & same & Relu & L2(0.05) \\ \hline 
MaxPool2D & 
&0&&\\ \hline 
Conv2D & (3,3)
& 4640 & same & Relu & L2(0.06) \\ \hline
Conv2D & (3,3) 
& 9248 & same & Relu & L2(0.06) \\ \hline

MaxPool2D & 
& 0 & & & \\ \hline 

Flatten & 
& 0 & & & \\ \hline 
Dense & 
& 240300 & & Relu & \\ \hline 
Dense & 
& 38528 & & Relu & \\ \hline 
Dense & 
& 8256 & & Relu & \\ \hline 
Dense & 
& 2080 & & Relu & \\ \hline 
Dense & 
& 165 & & Soft- & \\ 
& & & & max & \\ \hline 
\end{tabular}
\caption{Hyperparameters pertaining to the architecture of the CNN built to analyze the continuous-readout ZTF images. The input shape to the first layer is (20,20,1). \thesis{The MaxPool2D (which are set with stride 2) layers reduce the size of the input by selecting the brightest value in each LxL patch and act as noise reduction elements.} Regularization is used to minimize overfitting. L2($p$) corresponds to ridge regression, which penalizes the sum squares of weights to a power $p$, encouraging smaller, more evenly distributed weights. The total number of parameters is 303,633, all of which are trainable. }
\label{tab:hyperparams}
\end{table}

\begin{figure}
\includegraphics[width=\linewidth]{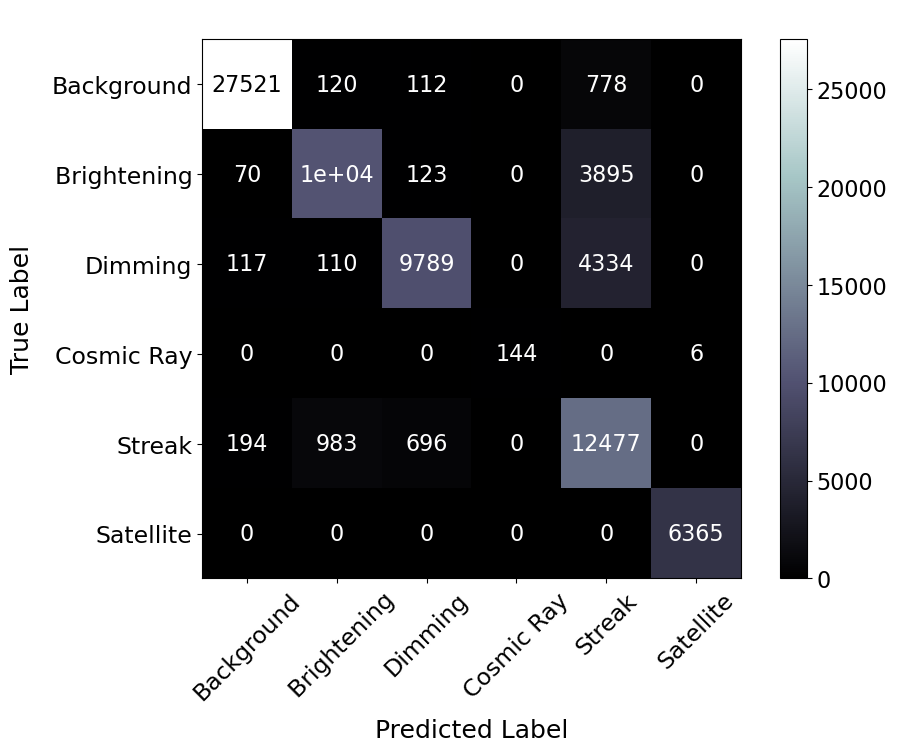}

\includegraphics[width=\linewidth]{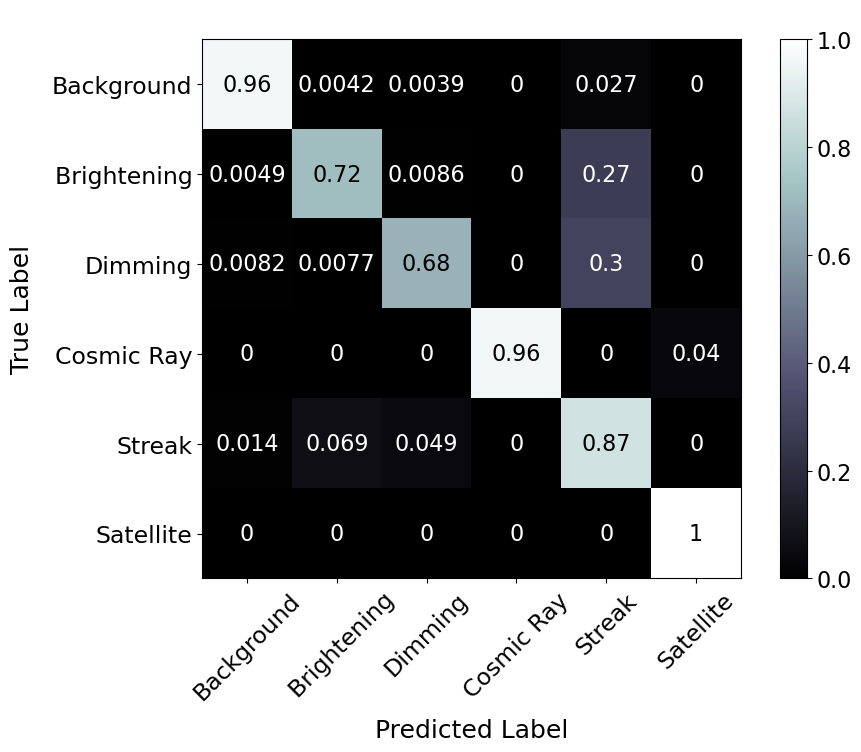}
  \caption{Confusion matrix for the CNN transients detection model. At this stage, the assigned class $C$ is set as the one with the highest probability $C = \vec{C}[argmax(\vec{P})]$.}
  \label{fig:confusion}
\end{figure}

\begin{figure*}
  \centering
  \includegraphics[width=\linewidth]{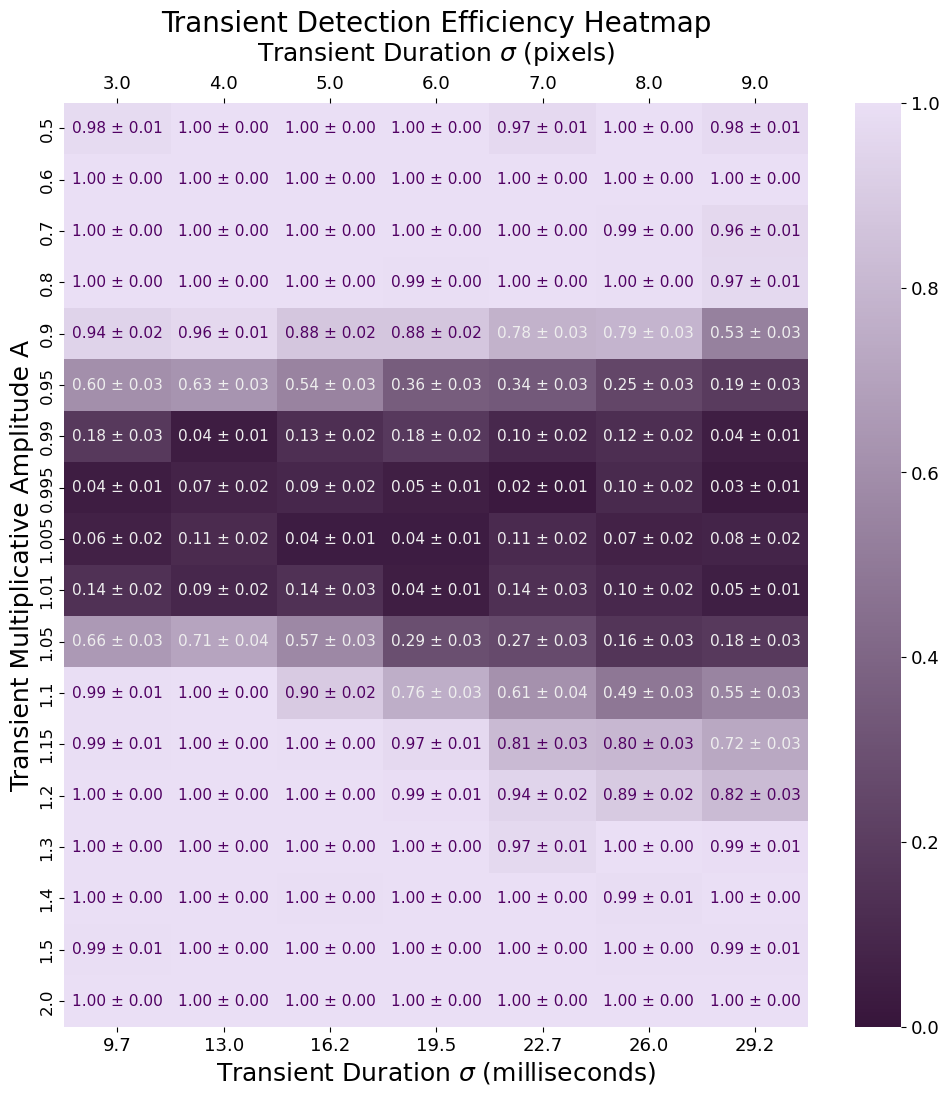}
  \caption{Retrieval efficiencies of the CNN  ($e_\mathrm{CNN}$) for implantations of different intensity (along the columns) and duration (along the rows). Efficiency is calculated as the number of correctly retrieved transients divided by the total number of implanted transients.  
  }
  \label{fig:eff}
\end{figure*}

\subsection{CNN Hyperparameters, Training, and Performance Evaluation}\label{sec:modeltt}

Our CNN model (described in \autoref{sec:cnn}) was trained on the stamps described in \autoref{sec:injections}. \thesis{In supervised learning, a subset of the data, the training dataset, is used to optimize model weights for the task, in this case, a multiclass classification. The remaining data, the test set, is used to produce predictions based on which the model performance is reported.} We reserved 25\% of the data for testing. We used a 5-fold cross-validation schema to measure the performance of the model during training and select the model hyperparameters (\eg{} duration of training). Unless otherwise specified, the CNN's performance is measured based on the average of the predictions of the five-fold model. \thesis{Neural networks are trained in successive epochs with specifically designed variants of the gradient descent algorithm to minimize an established loss function.} 
\autoref{tab:hyperparams} displays the details of each layer of our CNN. We set training to extend to 1,000 epochs for each validation set but used an ``early stopping callback'' that ended training when the validation loss didn't improve by more than 0.0001 
over 50 epochs and then restored the model to the weights that produced the best validation loss after training ended. 
We tested batch sizes of 10-500 stamps and found that a batch size of 50 stamps led to the best overall performance while reducing overfitting.

For this multiclass classification problem, we used a categorical cross-entropy loss and the Adam optimizer.
\thesis{The categorical cross-entropy loss function is:
$L(y,\hat{y})=- \sum_{i=1}^{N_c} y_i \log (\hat{y}_i)$, where $L(y,\hat{y})$ is the loss, $y_i$ is the true label for each class i, and $\hat{y}_i$ is the predicted probability for each class.} The learning rate \thesis{, the hyperparameter that controls how significantly the model's parameters are adjusted at each epoch of training, }was progressively reduced at 25, 50, 80, 110, and 150 epochs to be $1\times10^{-3}, 5\times10^{-4}, 1\times10^{-4}, 5\times10^{-5}, 1\times10^{-5}, 5\times10^{-6}$
respectively. \thesis{This allows the model to quickly find a better region of parameter space in the early stages of training, and then make fine adjustments in later stages.} 
The loss curves and CNN accuracy (fraction of correct classifications) as a function of the training epoch for each training and validation set are plotted in \autoref{fig:lossacc}. As discussed in \autoref{sec:cnn}, the output of our CNN is a 6D vector of probabilities $\vec{P}$ associated with the 6 classes (\eg{} $p_\mathrm{dimming}$ for a dimming transient). At this stage of the analysis, the predicted label corresponds to the class predicted with the highest probability: $C = \vec{C}[argmax(\vec{P})]$ (we will implement threshold on probabilities for transients later, see \autoref{sec:efficiency}).

\subsection{CNN Transient Retrieval Metrics}\label{sec:metrics}

Here we discuss the performance of our CNN with metrics measured on a test set, 25\% of the full dataset of stamps described in \autoref{sec:injections}, which was unseen during training. \thesis{As discussed in \autoref{sec:cnn}, the prediction of the CNN is a 5D vector of class probability $\vec{P}$ for the 5 $\vec{C}$ classes.}  Defining for this calculation a positive as a transient label, and a negative as any other pixel-level label, we use the following metrics of performance: {Precision} [$Pr = \frac{\mathrm{TP}}{\mathrm{TP}+\mathrm{FP}}$], { Recall} [$Re = \frac{\mathrm{TP}}{\mathrm{TP}+\mathrm{FN}}$ ], and their harmonic mean, the $F1$ score: 
$F1 = 2\frac{Pr \cdot Re}{Pr+Re},$
 where $TP$ is the number of correctly classified transient pixels (true positives), $FP$ is the number of non-transient pixels classified as transients (false positives), and $FN$ is the number of missed pixel detections (false negatives). \autoref{tab:f1} shows the CNN's $F1$ scores on the testing stamps. $F1$ scores are reported individually for every class. Precision and Recall will be discussed in \autoref{sec:efficiency}. 

 \begin{table}[h!]
\centering
\begin{tabular}{|c|c|}
\hline
 & \textbf{$F1$ Score  } \\ \hline

{Background}  &0.975 \\\hline
{Brightening Transient} & 0.795\\ \hline
{Dimming Transient}  &0.781 \\\hline

{Cosmic Ray}  &0.980 \\\hline

{Streak}  &0.696 \\\hline
{Satellite}  &0.9995 \\\hline

\end{tabular}
\caption{$F1$ scores for CNN detection with classification set to the highest probability class.}
\label{tab:f1}
\end{table}

\begin{figure}
\includegraphics[width=\linewidth]{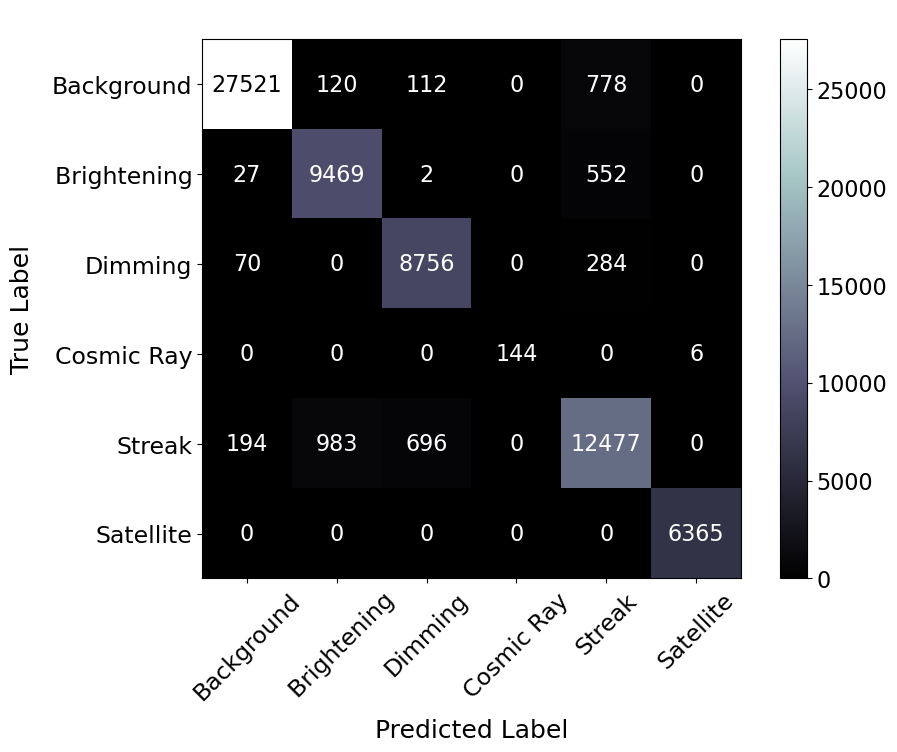}

\includegraphics[width=\linewidth]{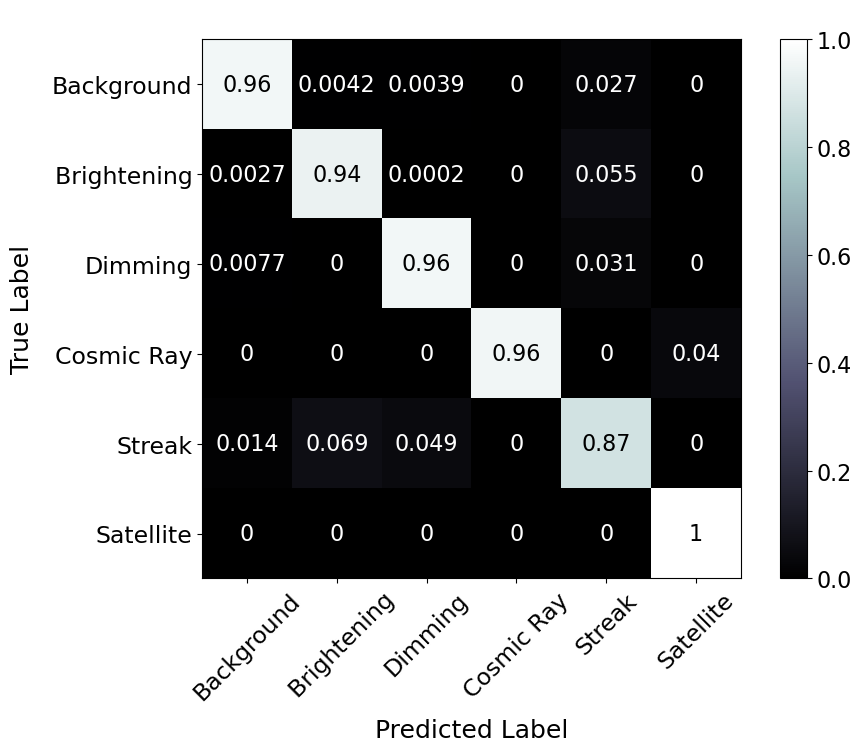}  \caption{Confusion matrix for the transients with an amplitude change of $\geq10 \%$. 
  }
  \label{fig:confusionsubset}
\end{figure}

\begin{figure*}[t!]
  \centering
  \includegraphics[width=12cm]{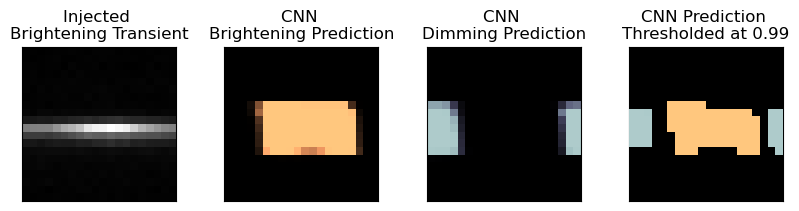}
  \caption{An example of the CNN's pixel-wise predictions, as probabilities in the two central panels and thresholded to make a True/False map in the rightmost panel. From this map, True pixels are then aggregated into a transient candidate for each type of transient (brightening and dimming). We note that pixels belonging to ingress and egress of a brightening transient are often classified as a dimming transient and vice versa.
}
  \label{fig:thrs}
\end{figure*}

\begin{figure}[h!]
  \centering
  \includegraphics[width=\linewidth]{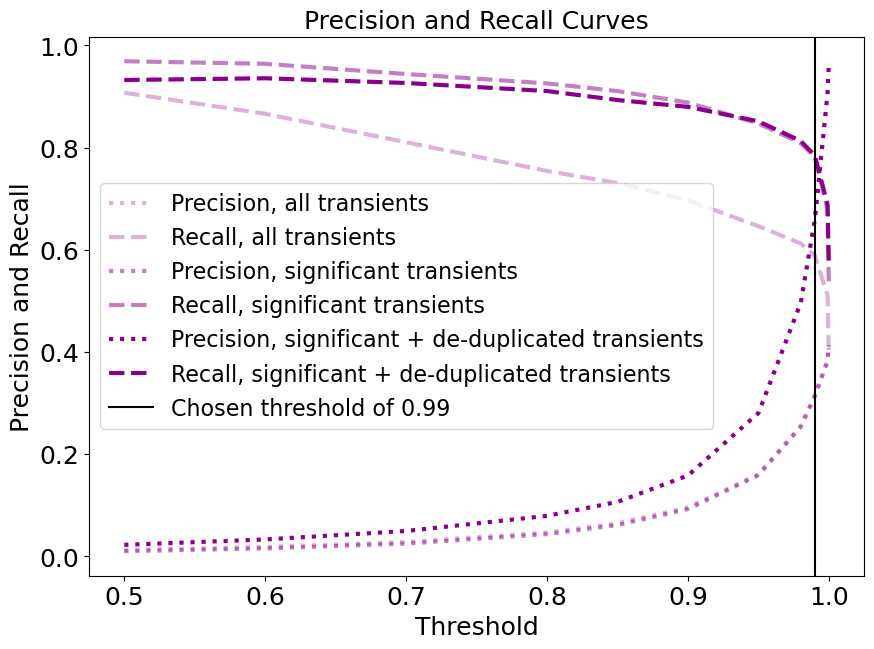}

  \caption{Precision and recall for the detection of fast transients with our CNN based on a varying probability threshold to identify transients. Precision and recall are reported for all transients; for significant transients with an amplitude change of $A \geq 10 \%$; and for significant transients that have been de-duplicated by aggregating proximate transient detections.}
  
  \label{fig:prec}
\end{figure}

\begin{figure*}[t!]
  \centering
  \includegraphics[width=\linewidth]{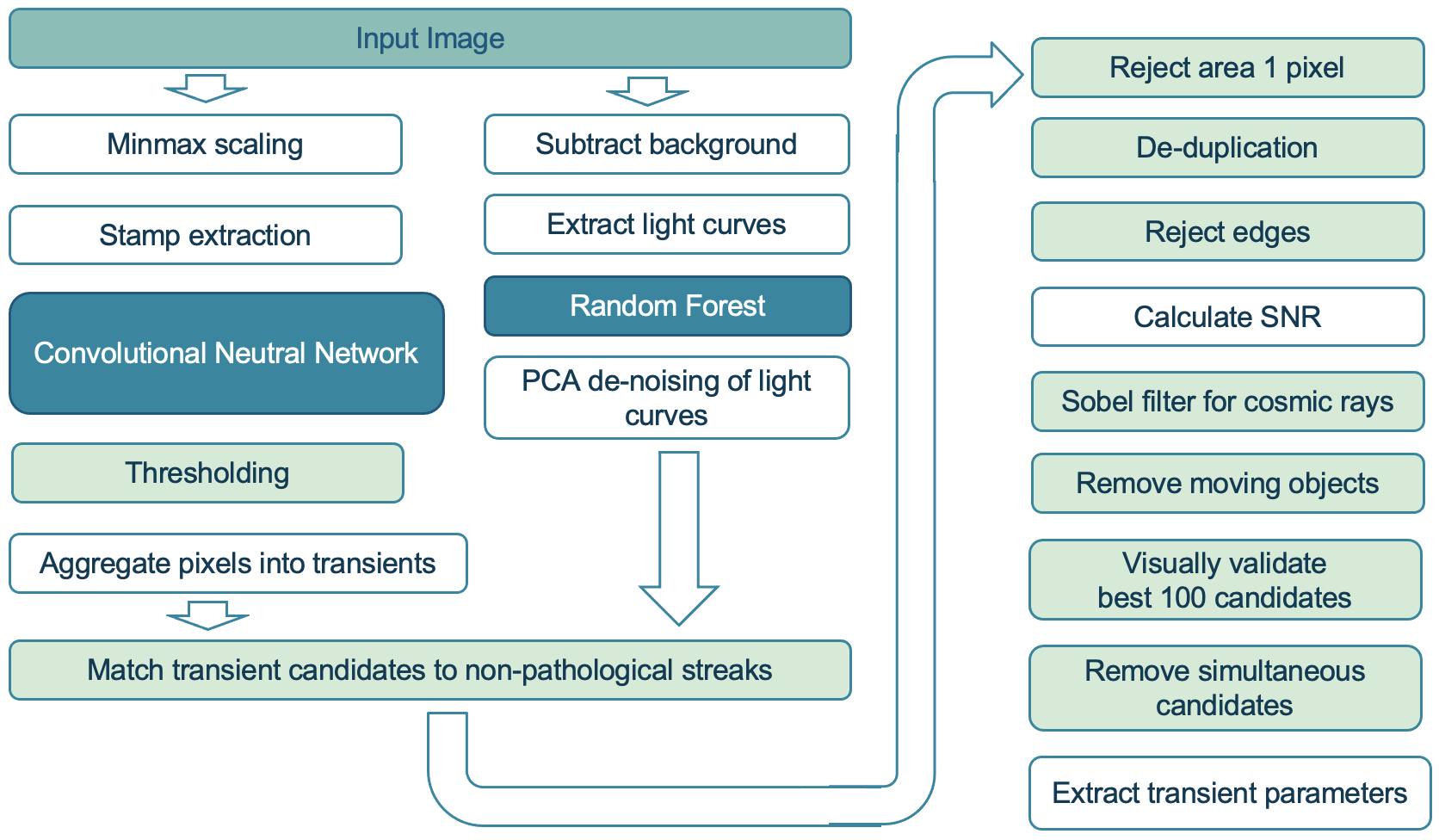}
  \caption{Processing pipeline. Steps that eliminate transient candidates are light green, and pure calculation steps are white.}
  \label{fig:processing}
\end{figure*}

The CNN performs at $F_1>0.97$ on background stamps, cosmic ray stamps, and satellite stamps. 
\autoref{fig:confusion} shows some confusion between brightening and dimming transients and between streaks and transients, consistent with the lower $F_1$ scores for these classes.

\autoref{fig:eff} shows how the transient amplitude and duration affect the CNN prediction. Efficiency drops when the transients have very small deviations from the original streak brightness but the CNN performs at $>$75\% efficiency for transients' amplitudes $A \geq 1.1$ and $A \leq 0.9$ with duration $w < 20$~ms, and detection is complete at $A \geq 1.3$ and $A \leq 0.8$. Longer duration transients are harder to find, which can be explained by our CNN seeing postage stamps of 20 pixels / 65 ms at a time. This suggests that different models (\eg{} transformer-based models) capable of ingesting full images) may be better suited to this task (Daniels et al. in prep.)

Confusion matrices reflecting the CNN's performance on \emph{significant} transients, defined as transients with $\geq 10 $ \% brightening or dimming, are shown in \autoref{fig:confusionsubset}. It is apparent that almost all of the misclassifications shown in \autoref{fig:confusion} can be attributed to transients with an amplitude change of $<$10\%, and the CNN can be trusted to categorize larger-amplitude events correctly, but its results may be incomplete or inaccurate at lower amplitude and this performance drop should be incorporated in any rate calculation.


\begin{figure}[h!]
  \centering
  \includegraphics[width=0.95\linewidth]{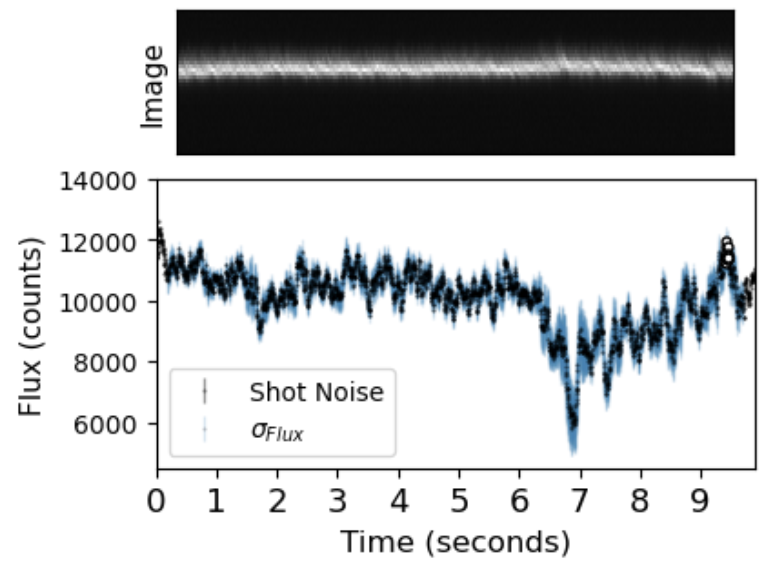}
  \includegraphics[width=0.95\linewidth]{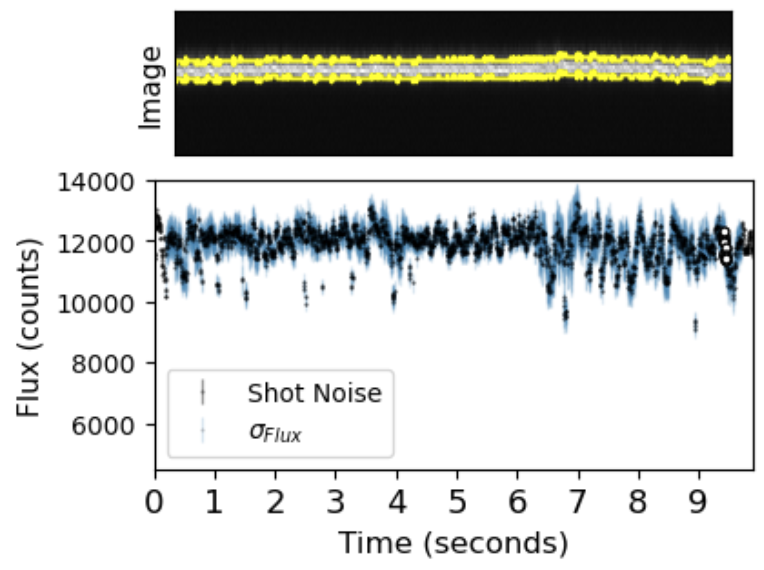}
    \caption{
    An example of a light curve with significant displacement along the spatial axis. The light curve is extracted without tracking (top panel) and with tracking (bottom panel). In each case, the top sub-plot shows the continuous-readout image; when tracking is applied, we show the edge of the tracking region in yellow. The bottom sub-plot shows the resulting extracted light curve. Uncertainties are marked as both $\sigma_\text{Flux}$ and $\epsilon_{Poisson}$ (Shot Noise in the legend, see \autoref{eq:ep}).
    With tracking, the flux baseline is rectified. Isolated flux drops appear in the tracking light curve; these may be genuine variability or may be introduced by the tracking process. We remove light curves where flux drops are artificially introduced by tracking at later stages of the pipeline (see \autoref{sec:randomforest}).
    }
\label{fig:tracking}
\end{figure}


\subsection{Event Retrieval and Efficiency}\label{sec:efficiency}

To retrieve transient detections from images, whether implanted for efficiency testing or to detect real transients, the pixelwise predictions must be aggregated into transient predictions. We aggregate adjacent transient pixels into a ``transient object'' by segmenting a Boolean detection array generated by thresholding the probabilistic output of the CNN (see \autoref{sec:efficiency}) 
using \texttt{scipy.ndimage.label} \citep{scipy} with 8-connectivity, or ``Moore Neighborhood''\citep{Sharma2013EdgeDU}. \thesis{That is, diagonally-adjacent True pixels were also considered to be part of the same transient (so for \texttt{scipy.ndimage.label}
the structure element was [[1,1,1], [1,1,1], [1,1,1]). This leaves us with an array of integers with each nonzero integer corresponding to a distinct transient candidate.} An example of this process is shown in \autoref{fig:thrs}. The center of each of these transient candidates is determined by calculating the geometric center for each integer in the label array, using \texttt{scipy.ndimage.center\_of\_mass} \citep{scipy}. Finally, the number of pixels in the transient candidate is recorded as the detection size.

Now we can shift our attention to the implementation of our model on full-scale images.
To test the CNN's prediction efficiency, 
we implanted transients in a set of full images, which we called our ``efficiency data." Five transients were implanted in the ten brightest streaks each for each quadrant of four representative fits files (1,600 implantations total): one from field 686 in the {\it i} filter, one from field 686 in the {\it g} filter, one from field 640 in the {\it i} filter, and one from field 449 in the {\it r} filter distributed evenly across all implanted amplitudes. We implanted a wider range of transient amplitudes in the efficiency data than in the training data to probe the limits of our CNN: in addition to the amplitudes listed in \autoref{tab:bright}, we included brightening amplitudes of $\times5$, $\times7$, and $\times10$, and dimming amplitudes of $\times0.4$, $\times0.3$, and $\times0.2$. 


We tested the CNN's predictions on the efficiency data 
for a variety of probability thresholds for the two transient classes after aggregation. Results are shown in \autoref{fig:prec}. 
Because our classification is based on $20\times20$ pixels postage stamps, many implanted transients are only partially included in a prediction image. As noted previously, the CNN classifies brightening and dimming transients separately. This can lead to pixels belonging to the ingress and egress of a brightening transient being classified as a dimming transient and vice versa. 
De-duplicating the transients by further aggregating detections within 30 pixels on the time axis allows us to remove several ``false positives" 
as shown in \autoref{fig:thrs}. A $p_{\mathrm{dimming/brightening}}> 0.99$ was chosen as the optimal balance of precision and recall (see \autoref{fig:prec}). The values of precision and recall at this threshold are shown in \autoref{tab:precision-recall}.



\begin{table}[h!]
\footnotesize
  \centering
  \begin{tabular}{|c|c|c|}
\hline
 & \textbf{Precision} & \textbf{Recall} \\ \hline
 {All transients} & 0.31 & 0.59 \\ \hline
{Significant transients}  & 0.31 & 0.78 \\\hline
{De-duplicated significant transients}  & 0.66 & 0.78 \\\hline
\end{tabular}

  \caption{Precision and recall for implanted transients, requiring a threshold of $p_{\mathrm{dimming/brightening}} \geq0.99$ to classify stamps as a brightening or dimming transient (see \autoref{fig:prec}).}
  \label{tab:precision-recall}
\end{table}

In this way, 5,197,200 pixels predicted by the CNN to be transients with p$>$0.99 are aggregated to give us 868,249 initial transient candidates.

\section{Post-CNN Processing}\label{sec:postcnn}
The ZTF continuous-readout mode complete data processing pipeline workflow is shown in \autoref{fig:processing}, and the steps following the CNN output are detailed in this section.

First, we extract light curves from the fits file.
Light curves are calculated and saved for the 25 brightest  sources in each image; see \autoref{sec:crossmatching}.

\subsection{Light Curve Extraction}\label{sec:lightcurveextraction}

To extract the light curves from the images, we use a 1-dimensional aperture on the spatial dimension across the width of the streak.
The instantaneous point spread function (PSF) of our sources is expected to be small, because the integration time for each row in our streaks is short ($\approx 0.003~sec$) diminishing the impact of tim integration over atmospheric wavefront \citep[\eg][]{2009ApJ...692..924L}. Nonetheless, the telescope contribution to the PSF is known to be $\approx 1.6''$ at the Palomar P48 telescope.

To determine the size of the aperture, we fit a 1D Gaussian along the spatial direction of an image for 100 streaks across 10 images. For the midpoint of each streak, the average standard deviation is measured to $<{\sigma_a}> = 0.99 \pm 0.8 ~\mathrm{pixels}$ (FWHM=2.33 pixels). Taking a similar calculation for the average of each streak results in $<{\sigma_a}> = 0.93\pm 0.1 ~\mathrm{pixels}$, and randomly sampling in a uniform distribution the time-point on each streak used for the calculation results in $<{\sigma_a}> = 0.98\pm 0.7 ~\mathrm{pixels}$. 
Therefore, we set the aperture $a=5$ pixels, corresponding to $\approx5$ standard deviations, or $\approx2.12\times \text{FWHM}$. We assume a similar PSF, and thus the same aperture, for all images within the single night of observing we analyze herein.

The centroid of the streak may fluctuate, for example due to vibrations in the telescope system or atmospheric effects. Because of this, we let the center of the aperture be the brightest pixel for each time step (within three pixels on either side of the detected transient candidate), thereby tracking the streak when it fluctuates in position. Quantifying the error in flux $e_F$ as the standard deviation $\sigma_F$ of the flux within the aperture $a$ over its mean $\mu_F$, averaged over ten light curve streaks in the same image ($e_F = \sum_{i=1...10} \frac{\sigma_{F,i}}{\mu_{F,i}}/10$), tracking reduces the error from $e_F = 0.11$ (with $\sigma_F= 0.049$) to $e_F = 0.033$ (with $\sigma_F= 0.003$). A visualization for this tracking procedure is provided in \autoref{fig:tracking}. For comparison, the shot noise for the flux, is calculated as 
\begin{equation}\label{eq:ep}
  \epsilon_{Poisson} = \sqrt{\frac{F}{G}+(N_\mathrm{read})^2}
\end{equation}
\noindent where $F$ is the pixel value in counts, $G$ is the gain, which for these observations is set to $G=6.2$, and the read noise is $N_\mathrm{read}=8.5$ for our data. Note that the shot noise is always lower than the standard deviation of the light curves, so the standard deviation is used in noise calculations (see \autoref{sec:snr}).



\subsection{PCA Denoising}\label{sec:pca}
A global denoising step is applied to the light curves to remove global trends. For each field, each CCD, and each observing run, light curves are concatenated along an observing sequence (typically 29 images, of which the first and last are excluded since they are incomplete in the readout mode). The sequences are aligned, low-pass filtered in Fourier space by applying a cutoff frequency of $3e^{-4}$ images or $\approx0.2$ seconds. Then a Principal Component Analysis (PCA) is performed \citep{jolliffe1986principal, pearson1901liii}. We find that typically $\approx10$ components reconstruct $>95\%$ of the variance in the light curve set. We thus defined the common trends as the PCA reconstruction of the set using 10 components and removed these common trends from the light curves. On a representative run (Field ID: 686, $g$ filter, CCD 07), this process reduces the variance of the median of all light curves by $\approx 35\%$.
\footnote{PCA decomposition is implemented the \texttt{sklearn.decomposition.PCA} \citep{scikit-learn} package and Fourier filtering via the \texttt{scipy.fft} package.}

\subsection{Machine Learning Rejection of Pathological Light Curves}\label{sec:randomforest}

\begin{figure}
\includegraphics[width=1.\linewidth]{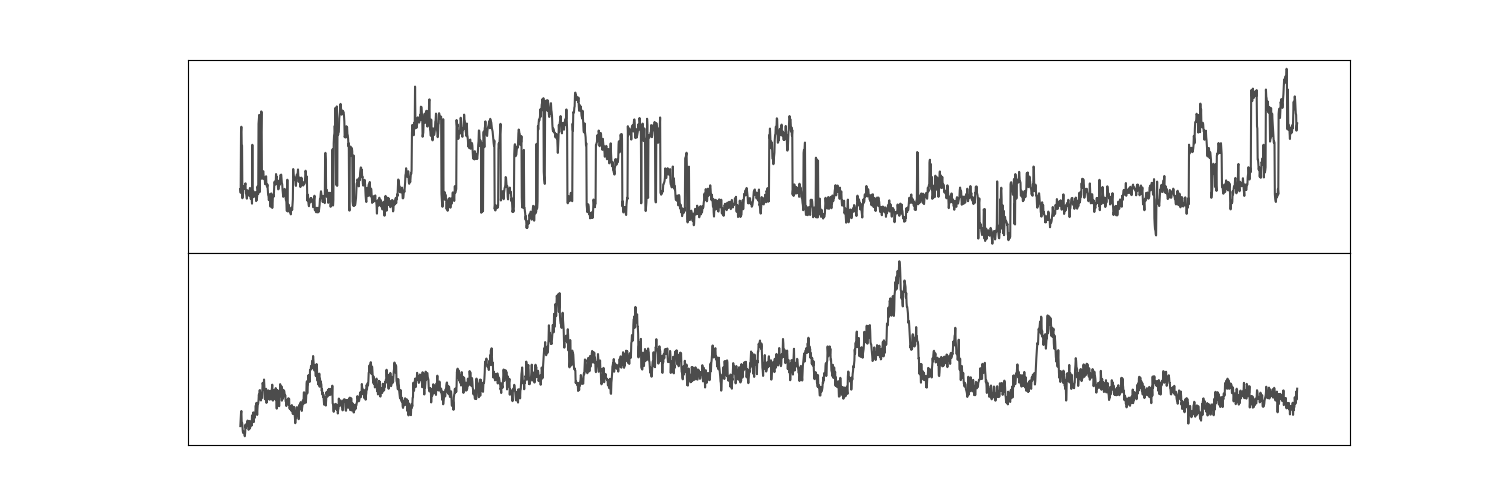}
  \caption{Examples of pathological light curves removed from the sample as transients in the relevant time scales could not be trusted due to the intrinsic light curve variability. Top: cross-contaminated flux between two light curve streaks entering the aperture. Bottom: high variability on the relevant time scale.}
  \label{fig:badlcv}
\end{figure}

We observed different phenomenology of pathological light curves. In particular, light curves where streaks from different sources merge and unmerge along the time axis cannot be reliably analyzed to identify transients (\autoref{fig:badlcv}, top). Our priority is to remove these light curves. In addition, light curves that show erratic variability at time scales relevant for the transients of interest in our work should also be removed from the sample (\autoref{fig:badlcv}, bottom), although the amplitude of variability in a light curve that makes it entirely unreliable is to some degree subjective. Tracking may also induce artifacts that may manifest as repeated flux drops (see \autoref{sec:lightcurveextraction} and \autoref{fig:tracking}). If left in the dataset, these light curves would lead to an overwhelming number of false positive detections that would need to be ruled out by inspection of the light curve itself. With nearly 800,000 light curves at this stage, we designed a machine learning step to reject ``pathological'' light curves. 
A small group of coauthors (2) visually inspected 734 light curves distributed across the different runs, filters, and fields and independently labeled them as ``retain'' or ``reject''. The team agreed on 
697 labels (538 to retain and 159 to reject). With these labels, we trained a Random Forest Classifier. 

The data was split by reserving 20\% of the light curves as test and performing a 5-fold cross-validation (CV) on the remaining 80\% with stratified splits to mitigate the class imbalance in the labels. The predictions of the 5 CV models are aggregated with weights corresponding to each fold's validation accuracy.

\begin{figure}
\includegraphics[width=1.\linewidth]{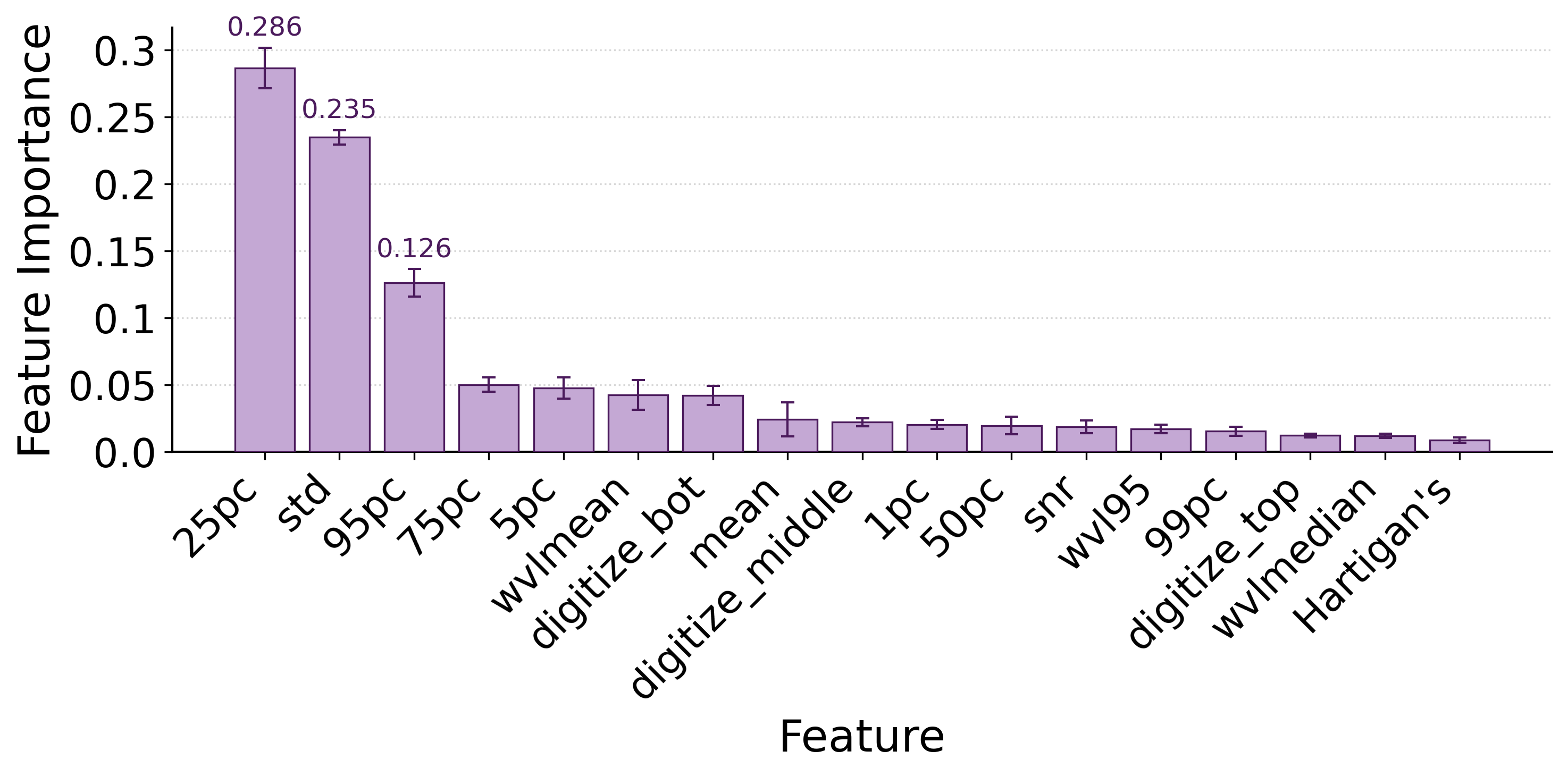}
  \caption{Aggregated Feature Importance from the 5-Fold Random Forest model that rejects pathological light curves. The labels are defined in \autoref{sec:randomforest}. The feature importance for the top three features is indicated in the plot.
  }
  \label{fig:featureimportance}
\end{figure}

The input to the random forest was 17 features extracted from the light curves: the mean, and standard deviation of the light curve flux (\texttt{std} in \autoref{fig:featureimportance}), as well as their ratio (\texttt{snr}); the following percentiles of the flux distribution: 1\%, 5\%, 25\%, 50\% (median), 75\%, 95\%, 99\%; the Hartigan's dip score \citep[\texttt{Hartigan's}][]{hartigan1985dip}, where the Hartigan's test for unimodality evaluates the null hypothesis of unimodality by measuring the maximum vertical discrepancy between the empirical Cumulative Distribution Function (CDF) and its closest unimodal fit, and a significant dip indicates multimodality--the Hartigan's dip score is the test statistic;\footnote{The implementation of the dip test is a modification of the code from \url{https://github.com/alimuldal/diptest}, modified for improved computational performance.} we perform a wavelet decomposition of the flux and include as features the mean (\texttt{wvlmean}), median (\texttt{wvlmedian}), and 95th percentile (\texttt{wvl95}) of the level‑1 detail coefficients (highest frequency components) from a 5‑level Haar wavelet decomposition of the low-pass filtered signal\footnote{The low-pass filtered signal is itself generated with wavelet denoising with a soft filter $S_{\lambda}(x) = \operatorname{sgn}(x) \cdot \max(|x| - \lambda,\; 0)$}; finally, noting that time series that show significant artifacts in their variability (and are thus to be rejected) have values that are much more dispersed around zero than those without such artifacts, we clip each time series to $\pm$10 and then minmax normalize it. We then created a 2-dimensional binning of the time series with 100$\times$100 bins, equivalent to a pixelated ``image'' of the time series in two dimensions. Time series that do not have significant variability are concentrated in the middle of this binned ``image'', and so we create three features for our random forest that are the total number of bins that contain light curve points in rows 40-60 in the binned image (\texttt{digitize\_middle}), above row 40 (\texttt{digitize\_top}), and below row 60 (\texttt{digitize\_bot}).

We use the \texttt{RandomForestClassifer} \texttt{Scikit Learn} implementation. Model hyperparameters are set to maximize performance and prevent overfitting: \texttt{n\_estimators}=100, \texttt{max\_depth}=3, and otherwise left as default.
We select a conservative threshold of $p=0.7$ for the probabilistic output of the classifier: the lowest threshold at which we have no false positives (rejection of light curves labeled to retain) in the unseen test set. This choice leads to the following performance (on 140 test light curves): 
mean accuracy: $0.931 \pm 0.001$, mean precision: $0.906 \pm 0.031$, mean recall: $0.811 \pm 0.023$, mean F1-Score: $0.856 \pm 0.019$. The model then rejects 9950 ($\sim1.3\%$) of the light curves in our dataset.
We also assessed whether the model may reject a light curve \emph{because} of the presence of a transient in it by implanting the light curves in the test set with exactly one Gaussian signal in each (randomly selecting different amplitudes and durations and including both positive and negative transients) and find that, on this modified set, the prediction for one out of 140 light curve changed, rejecting a light curve that we had labeled to retain. However, upon visual inspection, that light curve does have multiple ``drops'' which would make the detection of a dimming transient unreliable, and the implanted Gaussian has an extremely high amplitude. We conclude that the random forest carries a risk of $<1\%$ of rejecting detection.

\autoref{fig:featureimportance} shows the importance of each of the 17 features. The 95th percentile of the wavelet decomposition coefficient, variance, ratio of mean to variance, the digitized middle and top features, and the highest percentiles of the distribution (95th and 99th) are the most significant discriminators.

\subsection{Removal of Spurious Detections}\label{sec:cuts}

Starting with 868,249 transient candidates, we reject false positives in the following order: 
\begin{enumerate}
\item We discard transient candidates that have a center $\geq$5 pixels from the center of a streak (see \autoref{sec:crossmatching} for a description of the streak-finding process and \autoref{sec:randomforest} for our determination of viable streaks), eliminating 294,671 candidates;
\item We discard transient candidates with a size of one pixel, eliminating 169,192 candidates; in implanted data, this cut rejects the detections of 23 out of 956 implanted transients (efficiency $e_\mathrm{1pixel} \geq 0.976$);
\item We de-duplicate the transient detections by combining brightening and dimming detections as described in \autoref{sec:efficiency}, eliminating 75,922 candidates;
\item We reject transients near the edges: any transient candidate within 20 pixels of the edge of the image is removed, since the CNN was not trained to interpret the edges, as well as rejecting transient candidates that are within 100 pixels of the start of a continuous-readout series of images, as we find several artifacts in these regions. This step eliminates 8,587 candidates.
\end{enumerate}

We now have 317,140 remaining transient candidates.

\subsection{Signal to Noise Ratio Determination}\label{sec:snr}

\begin{figure}[t!]
  \centering
  \includegraphics[width=\linewidth]{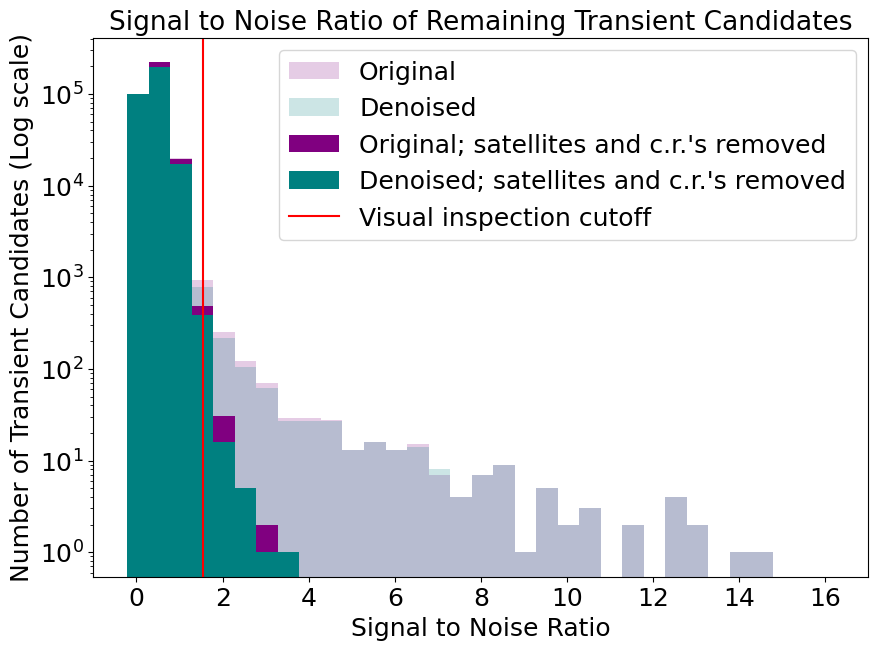}
  \caption{Signal to noise (\SNR ) ratio of transients that survived the cuts described in \autoref{sec:cuts}, plotted at high transparency, and ratio of those that survived the additional cuts to remove cosmic rays (c.r.'s) and satellites described in \autoref{sec:sobels}, plotted at full opacity. The local SNR distributions from both original and denoised light curves (\autoref{sec:snr}) are shown. Candidates above the red line are visually validated in \autoref{sec:candidates}. The histogram is cropped; three additional cosmic ray/satellites are located at SNR=31, 37, and 40 respectively.
}
\label{fig:SNRS}

\end{figure}

We measure the signal of an event as the area within the transient flux from the source's baseline flux, which we calculate in 
the following steps. Consider an event candidate tagged $k$ occurring in a given light curve. The light curve flux value for the $i-th$ point will be denoted as $f_{i}$. $f_{k}$ denotes the flux at the pixel tagged by the CNN.
\begin{itemize}
  \item{The brightest (dimmest) point in light curve $j$ within 10 points on either side of candidate $k$ is chosen as the central point: $$c=max(f_{i\in[k-10,k+10]}),$$ 
  $$c=min(f_{i\in[k-10,k+10]}),$$ for brightening and dimming transients respectively. Its flux is denoted as $f_{C,k}$. For aggregated transients, the location of the transient $k$ is chosen within $k_{min}-10,k_{max}+10$.}
  
  \item{The baseline flux $b_{k}$ is calculated as the average of the light curve flux 210-110 points on either side of the central point: $$b_{k} = \frac{\sum_{i\geq c-210}^{i < c-110}f_{i} + \sum_{i > c+110}^{i\leq c+210}f_{i}}{200}.$$} 
  \item{The amplitude of the event is calculated as $h_{k}=f_{C,k} - b_{k}$;}
  \item{The ``half-max'' of the event is calculated as $h_{0.5k} = 0.5h_{k}$. The integrated signal $S$ is the sum of flux relative to baseline for all points around $c$ until the first point with amplitude that drops below the half-max. $$S=\sum_{i: |f_i-b_{k}| > |h_{0.5k}|} {|f_i-b_{k}}|.$$}
  \item{The noise is calculated twice, on a local scale and a global scale. The local noise $N_l$ is the standard deviation of the points included in the baseline calculation: $$N_l=
  \sigma(f_{i\geq c-210 \& i < c-110} \cup$$ $$f_{i > c+110 \& i\leq c+210})$$ and the global noise is the standard deviation of all measurements farther than $110$ points away from the transient candidate's center: $$N_g=
  \sigma(f_{i <c-110} \cup f_{i > c+110})$$.}

\end{itemize}

The SNR is then $S/N_l$ (local) and $S/N_g$ (global).

The calculation is done on a high-pass filtered version of the PCA de-noised light curve (see \autoref{sec:pca}). The high-pass filtering is obtained by convolving the light curve with a Gaussian filter of standard deviation $\sigma_{\text{filter}}=27$ pixels (3 times the standard deviation of the longest implanted transients). 

The SNRs of the remaining transient candidates are shown in \autoref{fig:SNRS}.
  
\begin{figure}[t!]

  \includegraphics[width=0.9\linewidth]{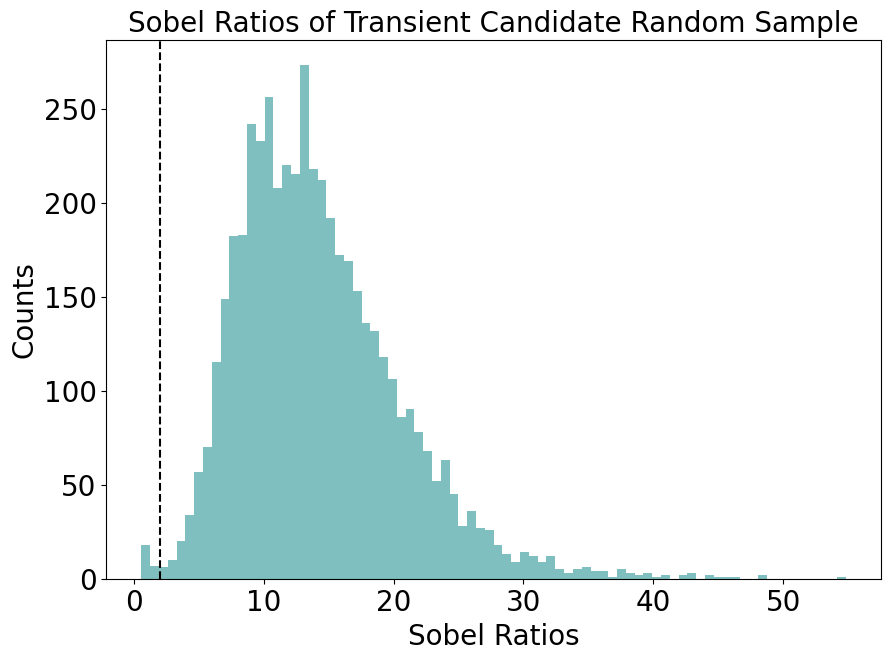}
  \includegraphics[width=0.9\linewidth]{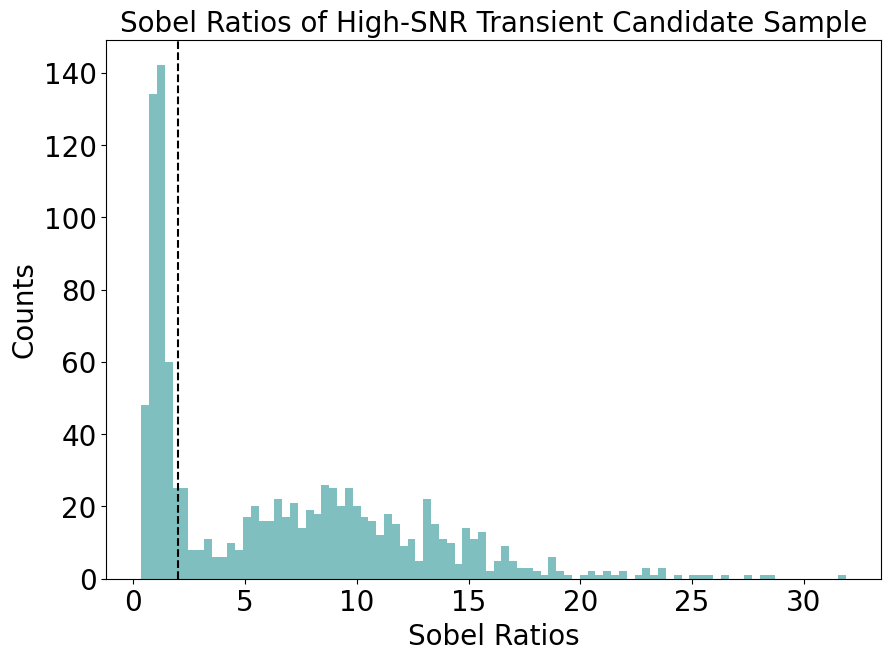}
  \caption{Sobel ratio ($S_r$) distribution for transients identified in our post-CNN analysis (see \autoref{sec:sobels}). The dashed line marks $S_r = 2$. The upper panel shows the Sobel ratio distribution of a random sample of transient candidates, and the lower panel shows the Sobel ratio distribution of brightening transient candidates with SNR$ >1$, where anything below the dotted line corresponds to the false positives that are removed at this stage in our pipeline.}
  \label{fig:sobels}
\end{figure}

\subsection{Additional Removal of Cosmic Rays and Satellites}\label{sec:sobels}

\begin{figure}[t!]
  \centering
  \includegraphics[width=\linewidth]{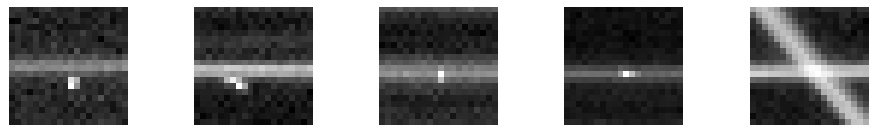}
  \caption{Examples of transient candidates that were removed with Sobel ratio $S_r \leq 2$. The first two panels show various cosmic ray morphologies; panels three and four show single-column brightening in each dimension, which is inconsistent with the PSF of a real source; and the final panel shows a satellite crossing.
}
\label{fig:sobel_examples}

\end{figure}
Visually inspecting our highest-SNR transient candidates, we find that we still have a significant contamination from cosmic rays and satellite crossings. To address the cosmic rays, we apply the Sobel–Feldman edge detection filter \citep{sobel2022sobel}, and calculate the ratio of edge sharpness along the $x$ and $y$ axis. We expect the transients to have smooth edges, especially along the time axis, while we expect cosmic rays to have sharp edges in both the spatial and temporal dimensions. We apply the Sobel filter, as implemented in \texttt{scipy.ndimage.sobel} \citep{scipy}, to the $10\times10$ -pixel region around the central point $c$ of each transient $k$ in each light curve (\autoref{sec:snr}). 
We define the Sobel ratio, $S_r$, as the ratio of the Sobel edge value at the boundaries of a transient object between the horizontal and vertical directions. A distribution of $S_r$ values for the transient objects at this stage of our analysis is shown in \autoref{fig:sobels}. We empirically find that cosmic rays in our data typically result in brightening events having a $S_r \leq 2$ and SNR $>1$.

To ensure that the removal of brightening events with $S_r \leq 2$ and SNR $>1$ does not remove short duration events ($\sim3~ms$, or a single row in our data), we visually inspected 400 of these events and found 13 events consistent with a single row brightening but morphologically ambiguous (\emph{\ie{}} not obviously a cosmic ray). We thus determine that the cut carries a risk of $<3.3\%$ of rejecting non-cosmic ray detections ($e_\mathrm{Sobel}\geq0.977)$.  However, we note that the PSF of our astrophysical sources is $\sim2$ pixels (\autoref{sec:lightcurveextraction}), which makes these events inconsistent with an origin above the atmosphere. In addition, among the 400 inspected events, we found 14 satellite crossings (which we removed from our data). Examples of transient candidates that were removed with this cut are shown in \autoref{fig:sobel_examples}. In total, this cut removes 1,167 transient candidates.

We chose a conservative threshold $S_r \leq 2$ so as to not remove any real transients, such there are still some remaining cosmic rays in our data --- but in a manageable number, which we can now remove by visual inspection.

We also note that, in spite of the high performance measures in our test data on CNN classification of satellites ($F1 = 0.9995$, \autoref{sec:metrics}), at this stage in our analysis, we found that the top SNR transient candidates are still contaminated by satellite crossings. To truly represent all satellites in the training, we would have to simulate a large number of satellite crossing angles and brightnesses. The satellites are, however, extremely apparent by eye, have high \SNR , and are limited in number. 
We thus manually remove 288 high SNR events corresponding to cosmic rays or satellites, leaving us with 315,685 candidates.

\begin{figure}[t!]
  \centering
  \includegraphics[width=\linewidth]{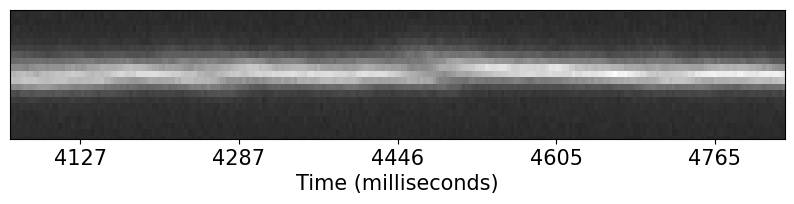}
  \caption{A broadening of the PSF, pictured in the center of this image. Less significant broadenings are visible on either side.
}
\label{fig:smudges}
\end{figure}

\section{Validation of Retrieved Transient Candidates and Transient Rate Limits}\label{sec:candidates}

The 100 highest-SNR remaining transient candidates were visually inspected and validated, beginning with a check for simultaneity. The minimum SNR of these candidates was 1.55 (see \autoref{fig:SNRS}).

\subsection{Removal of Simultaneous Candidates}\label{sec:simult}

True astrophysical phenomena will not affect more than one streak simultaneously. Recall that our streaks are not aligned in time due to offsets in the initial $x$ position of the sources on the CCD. Transient candidates in the same CCD time-step (column), or ``simultaneous in position," may be due to instrumental effects. Candidates that are ``simultaneous in time," where accounting for the offset in the initial position of the sources on the CCD leads to simultaneity, may be due to instrumental effects or caused by atmospheric phenomena (changes in seeing, clouds etc...). For our top 100 candidates, we check for simultaneous candidates against all other transient candidate with SNR $>$ 0.1. If any candidates are flagged as simultaneous to within 20 pixels ($\sim65$ ms), we inspect the light curves of matching events as well as the light curves of the other nearest streaks in the image, to assess whether simultaneous transient events with similar phenomenology truly existed (locally on the image or globally across it). Of our highest-SNR candidates, one candidate was found to be simultaneous in time and removed from consideration, and no candidates were found to be simultaneous in position.

\subsection{Visual Validation of Transient Candidates}

\begin{figure}[t!]
  \centering
  \includegraphics[width=\linewidth]{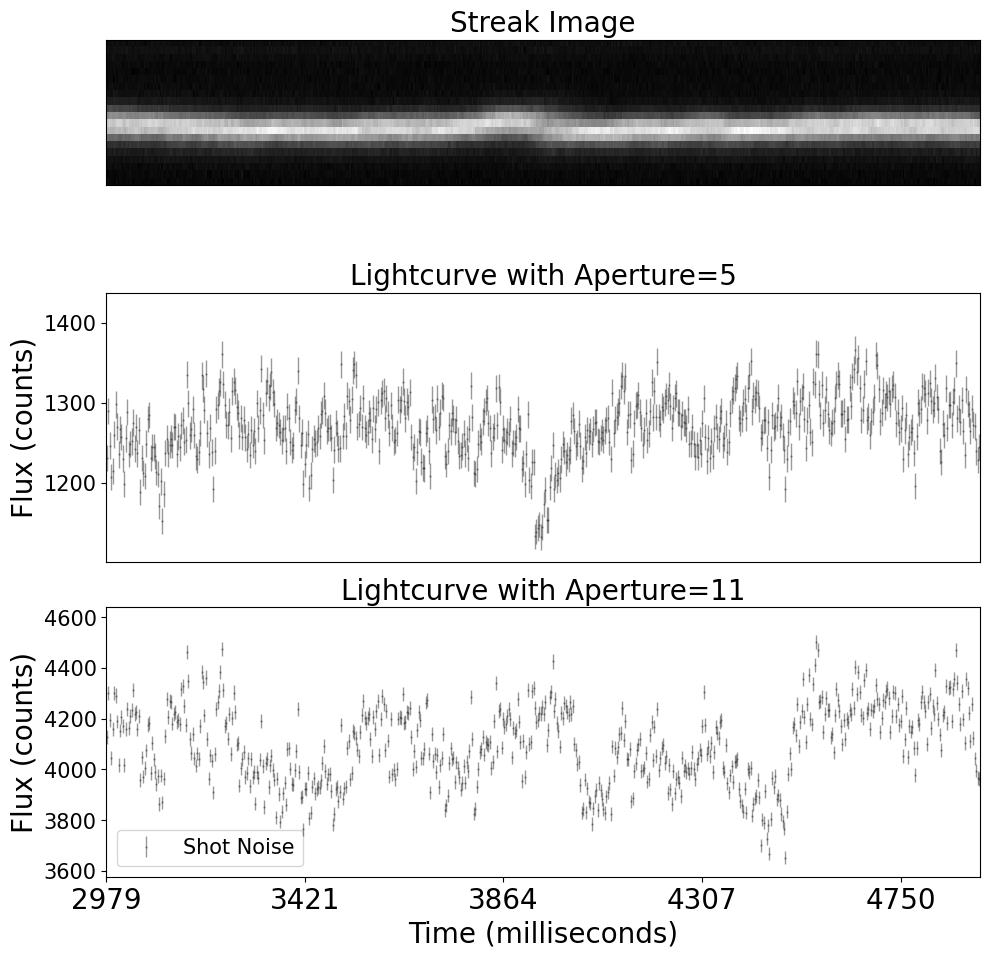}
  \caption{An example of a transient candidate where a dip is visible in the 5-pixel aperture light curve but is only as significant as the rest of the noise in the 11-pixel aperture light curve. These events are typically due to broadening or shifting of the flux along the spatial position and could be due to atmospheric effects or telescope vibrations. Events due to telescope vibrations, however, would be simultaneous in time and therefore would be removed in earlier steps. Atmospheric events may be localized.
}
\label{fig:aperturetest}
\end{figure}
The remaining 99 candidates were visually validated: the filtered and unfiltered light curves, the light curve obtained from a larger aperture of size 11 pixels, and the full image and image cutout were all inspected.

The authors conclude that, while the pipeline did find many instances where streaks show deviations from their typical behavior, none of the phenomena are consistent with astrophysical transients. A breakdown is as follows:

\begin{figure*}[t!]
  \centering
  \includegraphics[width=0.8\linewidth]{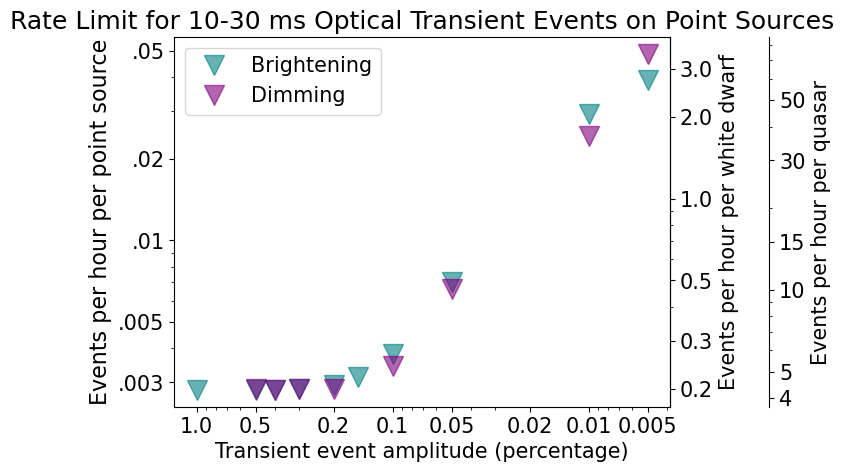} %
  \caption{Rate limits for optical brightening and dimming transient events with $\SNR{} \geq 1.55$ on $G_{RP}\leq 12$ point sources, as well as rate limits for events on quasars and white dwarfs (right-hand axes), based on nondetections in our data.
}
\label{fig:rate_limits}
\end{figure*}

\begin{itemize}
   \item{7 transient candidates are on streaks contaminated by the flux of nearby streaks. While we aimed at removing these light curves with the random forest (\autoref{sec:randomforest}) we had deliberately set a conservative cut for rejection, so it is not surprising that some of these ``pathological'' light curves would leak into the accepted dataset;}
   \item{25 transient candidates that have a high local SNR appear in noisy light curves where similar or larger flux deviations are common throughout;}
   \item 67 transient candidates were associated with atmospheric or scintillation effects. This manifests as either a broadening of the PSF or a lateral movement that was not fully captured by the tracking algorithm described in \autoref{sec:lightcurveextraction}. 
   In most case, these candidates are not present in the light curve extracted with a larger aperture. An example is shown in \autoref{fig:aperturetest} (since they are not simultaneous in position or time we rule out camera instability or vibrations of the telescope). In some cases, the broadening of the PSF is so significant that it still appears in an 11-pixel aperture light curve, but an inspection of the image shows that the flux is dispersed but still present. An example is shown in \autoref{fig:smudges}. One phenomena that operates on these timescales is lensing by a cold air bubbles in the upper atmosphere, which is expected to cause relatively localized flux dispersions. Lensing via a cold-air bubble would also explain the diagonal slant to the broadening shown in  \autoref{fig:smudges}, as the bubble travels across the telescope beam.

\end{itemize}


A method to retrieve features from detected transients, which is described and assessed in \autoref{sec:fitting}, but since no transients of astrophysical nature were identified, we do not report the parameters of any transients here.

\subsection{Transient Event Rate Limits}\label{sec:rate limits}

With the nondetection of astrophysical transients in our data, we can set a rate limit on optical transients of duration 10-30 ms associated with point sources. Our rate limits are shown in \autoref{fig:rate_limits}.

Combining the efficiency at different steps of our pipeline (which is determined as a function of amplitude $A$ for the CNN detections) as  $e_\mathrm{total}(A) = e_\mathrm{CNN(A)}*e_\mathrm{1pixel}*e_\mathrm{Sobel}$, and assuming Poisson statistics and a 95\% confidence interval, we define our per-source event rate as:

$$R_{95}(A)\leq\frac{2.996}{e_\mathrm{total} * h_\mathrm{ps}}{ h^{-1}},$$

\noindent where $h_\mathrm{ps}$ is the total observing time for all point sources of magnitude $G_{RP}\leq 12$. At fractional amplitudes $A=[0.2, 0.05, 0.005]$ our rate upper limits are $R_{95}(A)\leq[0.003, 0.007, 0.039] ~h^{-1}$ for brightening and $R_{95}(A)\leq[0.003, 0.007, 0.049] ~h^{-1}$ for dimming events at significance $\SNR\geq1.55$. Rates were also calculated for separately for white dwarfs (WDs) and quasars (QSOs) based on their respective number of observing hours estimated by crossmatching a fraction of the sources with Gaia. White dwarfs and quasars are particularly relevant in our inference since they are expected to show short-term variability \citep{burdge20208, marsh2016radio, aranzana2018short, otero2024optical}. For the same amplitudes, we report rate upper limits of  $R_\mathrm{95,WD}(A) \leq [0.207, 0.494, 2.74]~h^{-1}$ for brightening and $R_\mathrm{95,WD}(A)\leq[0.199, 0.464, 3.41]~h^{-1}$ for dimming events in white dwarfs and $R_\mathrm{95,QSO}(A) \leq[4.49, 10.7, 59.3]~h^{-1}$ for brightening and $R_\mathrm{95,QSO}(A) \leq [4.32,10.1,74.0] ~h^{-1}$ for dimming events in quasars.
These limits are shown in \autoref{fig:rate_limits}. To our knowledge, these are the first rate for optical events in the $\sim10$~ms regime in association with point sources to appear in the literature.

\section{Conclusions and Future Work}

We conducted a pilot survey in continuous-readout mode as a special survey within ZTF, developing an end-to-end analysis pipeline to identify millisecond transients in this unique observational mode. Our pipeline identified statistically significant flux variations whose origins are potentially astrophysical but also can be produced by contaminants, primarily in the form of cosmic rays and atmospheric events. This analysis enabled us to assess the prevalence and features of such contaminants in our data and design strategies for photometry which, due to the nature of the continuous-readout data, require custom-built analysis tools. With these data, we set the first systematic observational rate limits for 10-30 ms optical transient events on point sources in the optical night sky.

We analyzed in total \DATAVOL  of continuous-readout images and \SHanallyzed{} of data from approximately \Nstarsanalyzed{} stars, \Nquasarsanalyzed{} quasars, and \NWDanalyzed{} white dwarfs, estimated by crossmatching a fraction of the sources with Gaia.

The initial transient detection is performed on the raw continuous-readout images, in which sources have a unique ``streak'' morphology, with a CNN\footnote{We note that our patch-based CNN approach uses information local (set by the postage stamp size) to a given point in the image to use as a basis for pixel classification.  Transformer neural networks \citep{vaswani2017attention} applied to images \citep{han2022survey} would be a naturally good option for future analysis of these data since these architectures incorporate information across the whole image, potentially enabling predictions based on longer trends in each streak and the relative behavior of multiple streaks.} that reaches a  97\% efficiency for transients of amplitude $A\geq1.3$ and $A\leq0.8$. Detection completeness drops when the transients have very small deviations from the original streak brightness; the efficiency is  $>$72\% for transient amplitudes $A>1.1$ and $A<0.9$. Aggregating and thresholding the CNN predictions leads to a precision of $Pr=0.66$ and recall of $Re=0.78$ for transients of amplitude $A\geq1.1$ and $A\leq0.9$ (see \autoref{fig:prec} and \autoref{tab:precision-recall} and \autoref{sec:cnn_everything}). 

The efficiency of our subsequent light curve-level analysis and cosmic ray rejection was approximately 94.4\%: $\geq97.6\%$ in the rejection of 1-pixel CNN candidates (\autoref{sec:cuts}) and $>96.7\%$ from a Sobel ratio-based rejection (\autoref{sec:sobels}). The total number of candidates retrieved is 315,685 (\autoref{sec:postcnn}). Of those, we visually validated the light curves and images for the 100 highest-SNR detections ($\SNR\geq1.55$, see \autoref{sec:candidates}) and determined that they were not astrophysical in nature.

Thus, with this volume of data – and given the unique capabilities of continuous-readout mode – we derive the following rate limits for brightening and dimming events on point sources: $R_{95}(A)\leq[0.003, 0.007, 0.039] ~ ([0.003, 0.007, 0.049]) $ events per hour per source for brightening (dimming) events for amplitudes $A=[20\%, 5\%, .5\%]$ (\autoref{sec:rate limits}, \autoref{fig:rate_limits}).
Splitting our data by the nature of the source, this corresponds to an upper limit of $R_\mathrm{95,WD}(A) \leq [0.207, 0.494, 2.74]~([0.199, 0.464, 3.41])$ brightening (dimming) events per hour for white dwarfs and $R_\mathrm{95,QSO}(A) \leq [4.49, 10.7, 59.3] ~ ([4.32,10.1,74.0])$ brightening (dimming) events per hour for quasars.

To our knowledge,
these are the first survey-derived rate limits on transient events in the $10-30$~ms duration range in the optical regime.  While rapid photometry surveys have been conducted in the past \citep{ dhillon2007ultracam, 
2008AJ....135.1039B, 2006A&A...446..739L, 2012SASS...31..147G,arimatsu2017organized, arimatsu2021detectability, huang2021taos, sako2018tomo, zhang2024optical, nir2021weizmann, pass2017pipeline, howell1986time}, 
including using continuous-readout mode \citep{bianco2009search}, they have been at generally slower time scales (e.g. KBO occultations, $
\sim100$ ms), or on field transients not associated with bright sources \citep[\eg{} FRB counterparts][]{tingay2021high}. Thus, no prior results are directly comparable with ours.

This first quantitative characterization of the bright millisecond optical sky via an untargeted, wide-field survey serves as a pathfinder for the Argus array by establishing constraints on the optical millisecond transient background. Our rate limits confirm the rarity of high-significance optical phenomena at millisecond time scales while also demonstrating the promise of continuous readout for fast transient searches.

\section{Software}

\texttt{NumPy} \citep{harris2020array}

\texttt{Pandas}\citep{mckinney2010data}

\texttt{Matplotlib} \citep{4160265}

\texttt{SciPy} \citep{virtanen2020scipy};

\texttt{Astropy} \citep{robitaille2013astropy};

\texttt{scikit-learn} \citep{scikit-learn};

\texttt{TensorFlow} \citep{tensorflow2015-whitepaper};

\texttt{Keras} \citep{chollet2015keras};



\texttt{tqdm} \citep{da2019tqdm};

\texttt{diptest} \citep{maechler2013package};

\texttt{PyWavelets} \citep{lee2019pywavelets};

\texttt{seaborn} \citep{Waskom2021};

\texttt{plotly} \citep{plotly}.

\section{Acknowledgments}

The authors thank Prof. Maryam Modjaz for sharing her expertise on infant supernova for \autoref{fig:phasespace}.

The authors acknowledge and thank Prof. Shri Kulkarni and Prof. Mansi Kasliwal, P.I.'s of the ZTF project, for their vision in the creation of the ZTF project and support of this work.

This work is supported by an Amazon Web Services grant (PI Andreoni) and by the University of Delaware DARWIN computing system: DARWIN – A Resource for Computational and Data-intensive Research at the University of Delaware and in the Delaware Region, which is supported by NSF under Grant Number: 1919839, Rudolf Eigenmann, Benjamin E. Bagozzi, Arthi Jayaraman, William Totten, and Cathy H. Wu, University of Delaware, 2021, URL: https://udspace.udel.edu/handle/19716/29071.

FBB is partially supported by NSF AST Awards: 2511639 and 2308016, 2123264.

This work is based on observations obtained with the Samuel Oschin Telescope 48 inch at the Palomar Observatory as part of the Zwicky Transient Facility project.  ZTF
is supported by the National Science Foundation under grant Nos.
AST-1440341, AST-2034437, and currently Award 2407588.
ZTF receives additional funding from the ZTF partnership.
Current members include Caltech, USA; Caltech/IPAC, USA;
University of Maryland, USA; University of California, Berkeley,
USA; University of Wisconsin at Milwaukee, USA; Cornell
University, USA; Drexel University, USA; University of North
Carolina at Chapel Hill, USA; Institute of Science and
Technology, Austria; National Central University, Taiwan, and
OKC, University of Stockholm, Sweden. Operations are
conducted by Caltech’s Optical Observatory (COO), Caltech/
IPAC, and the University of Washington at Seattle, USA.

This material is based upon work supported by the National Science Foundation Graduate Research Fellowship Program under Grant No. 2444111. Any opinions, findings, and conclusions or recommendations expressed in this material are those of the author(s) and do not necessarily reflect the views of the National Science Foundation.

\appendix

\section{Transient Candidate Parameter Extraction through Template Fitting}\label{sec:fitting}

Although none of our transient candidates passed our inspection, we developed a fitting routine that would characterize retrieved transients. While different physical phenomena (\eg{} occultations, FRBs, Blazars) will manifest with different light curve morphologies (some of which are known, others are unknown), in this initial phase we use a general and simple Gaussian model, with the same structure as \autoref{eq:gaussian}:

\begin{equation}
  F_\mathrm{transient, i} = F_0\left((A-1)e^\frac{ -(y_i-\mu)^2}{2w^2} +1\right),
  \label{eq:gaussianfit}
\end{equation}

\noindent where $F_0$ is the source's baseline flux and $y_i$ the position of the $i$-th pixel along the temporal axis.
This template is fit to the flux by minimizing the sum of the absolute value of the difference between the real flux and calculated flux over 390 ms centered at the transient's geometric center (see \autoref{sec:efficiency}). The parameters $A$, $\mu$, $w$, and $F_0$ are all fit to the data.

Guesses for the initial values of the parameters are made as follows:
\begin{itemize}
  \item $\mu$ is initialized as the center of the transient as retrieved by the CNN,
  \item $F_0$ is initialized as the average of 50 pixels on each side of the detection, excluding the 10 closest pixels on each side of the detection,
  \item $A$ is initialized as the value of the flux at $\mu$ divided by the $F_0$ initial guess,
  \item $w$ is initialized as the square root of the number of pixels included in the transient detection as reported by the CNN.
\end{itemize}

\begin{figure}[t!]
  \centering
  \includegraphics[width=8cm]{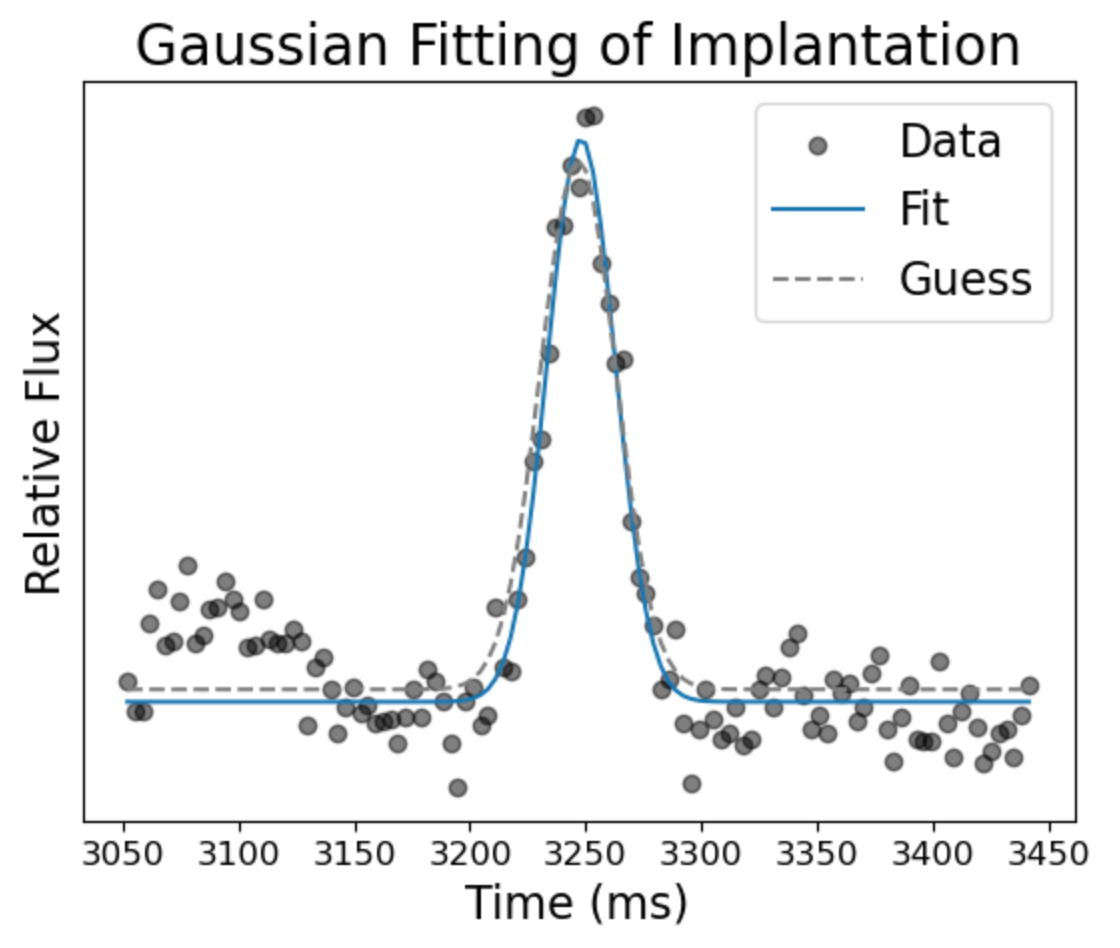}
  \caption{Example of template fitting for an implanted transient. In this example, the true values were: $\mu=3247$, $w=16.2$ ms, and $A=1.3$. The gaussian fitting retrieved: $\mu=3247$, $w=14.6$ ms, and $A=1.47$.
}
\label{fig:fitting}
\end{figure}

This fitting routine was tested on implanted transients. \autoref{fig:fitting} shows a visual example of the fitting on an implanted brightening event, and \autoref{tab:paramfitting} shows the results of the fit on implanted transients.

\begin{table*}[t!]
\centering
\begin{tabular}{|c|c|c|c|c|c|c|}
\hline
\textbf{$A$} & \textbf{$N_r$}& \textbf{$<\epsilon_A>$}& \textbf{$<\epsilon_\mu>$}& \textbf{$< \epsilon_w>$}&
$<|\mu_t - \mu_p|>$ & $<|w_t - w_p|>$\\ 

\hline

0.8 & 100 & 0.24 & 6.65e-03 & 5.25 & 6.48 & 4.2 \\
\hline
0.85 & 100 & 0.242 & 6.54e-03 & 4.94 & 5.67 & 4.2 \\
\hline
0.9 & 66 & 0.108 & 5.94e-03 & 3.73 & 4.96 & 3.36 \\
\hline
0.95 & 28 & 0.146 & 1.06e-02 & 5.04 & 8.29 & 4.79 \\
\hline
0.99 & 26 & 0.312 & 1.77e-02 & 5.6 & 12.5 & 5.55 \\
\hline
0.995 & 29 & 0.56 & 1.76e-02 & 4.16 & 11.8 & 4.14 \\
\hline
1.05 & 5 & 2.93e-02 & 1.00e-03 & 0.89 & 0.916 & 0.934 \\
\hline
1.1 & 20 & 2.59e-02 & 2.22e-03 & 2.76 & 2.44 & 3.04 \\
\hline
1.3 & 100 & 2.36e-02 & 8.74e-04 & 0.35 & 0.491 & 0.455 \\
\hline
1.5 & 100 & 2.49e-02 & 3.88e-04 & 0.251 & 0.359 & 0.377 \\
\hline
2 & 100 & 4.96e-02 & 3.94e-04 & 0.407 & 0.369 & 0.814 \\
\hline
3 & 100 & 3.70e-02 & 1.92e-04 & 6.90e-02 & 0.188 & 0.207 \\
\hline
4 & 100 & 6.12e-02 & 2.02e-04 & 7.57e-02 & 0.197 & 0.303 \\
\hline
6 & 100 & 3.53e-02 & 1.51e-04 & 1.33e-02 & 9.85e-02 & 7.97e-02 \\
\hline
8 & 100 & 3.63e-02 & 1.24e-04 & 1.41e-02 & 0.113 & 0.113 \\
\hline
10 & 100 & 3.43e-02 & 1.41e-04 & 6.53e-03 & 8.40e-02 & 6.53e-02 \\
\hline
\end{tabular}
\caption{Transients parameters as derived from fitting a Gaussian model to Gaussian implantations. 180 transients were implanted at each amplitude $A$. These statistics were derived on all transients recovered $N_r$ at a given $A$ or 100 events, whichever is larger. Errors are reported as absolute errors in units of pixels for $w$ and $\mu$ and fractional differences for all quantities, \eg{} $<\epsilon_A> =  \sum_i{\frac{|A_{t,i} - A_{p,i}|}{A_{t,i}}} / N_r$ where $i$ denotes the $i$-th transient and $t$ and $p$ denote the true and predicted values respectively.
The position $\mu$ (column number) and amplitude A (see \autoref{eq:gaussian}) are recovered with much higher accuracy than the duration of the transient $w$. Errors on $w$ are larger for dimming transients. }
\label{tab:paramfitting}
\end{table*}

\section{Crossmatching and streak identification}\label{sec:crossmatching}

Identifying the streaks as known stars/quasars/white dwarfs is not a trivial task due to the loss of information of one spatial dimension. Moreover, the WCS positions for this format of data was not reliable, so additional steps were taken to retrieve identifications.

To begin, images from the beginning of continuous-readout mode operation were analyzed to find the initial position of the streaks (see the left panel of \autoref{fig:combined}). We take the rolling mean along each column (spatial axis)
and record the row (time axis) position when it changes by $>$70 counts over a different of $\sim100$ pixels. The position of the greatest change is saved. Pixels separated by $<$8 pixels are considered to be part of the same streak. In the case of overlapping streaks, only the brighter one is retained, because the inherent variability of some streaks is quite high, so allowing more than one position along each streak meant that there were too many streaks which were not from different sources. This method cannot detect sources where the initial position is close to the edge of the CCD; in a sample of 108 images, 2.9$\%$ of streaks that were detectable through a peak-finding algorithm of a mid-CRM image did not have detected initial positions in the beginning-CRM image.

With the initial positions of the brightest sources in each field, we run Astrometry.net's solve-field with an XY list file \citep{lang2010astrometry}. Initial guesses for the pixel scale, RA, and Dec are taken from the original unreliable FITS header. Fields were solved with index 4107, or, occasionally, index 4108.

With WCS coordinates corrected to within a few pixels (uncertainty from the sources' initial positions), Gaia DR3 was queried to obtain identifiers for the streaks. Since we chose only the ten brightest streaks in each image for our training data, the brightest Gaia object (in the Gaia band closest to the ZTF band) within 9 arcseconds was matched to the streak. 

\begin{figure}[h!]
  \gridline{\fig{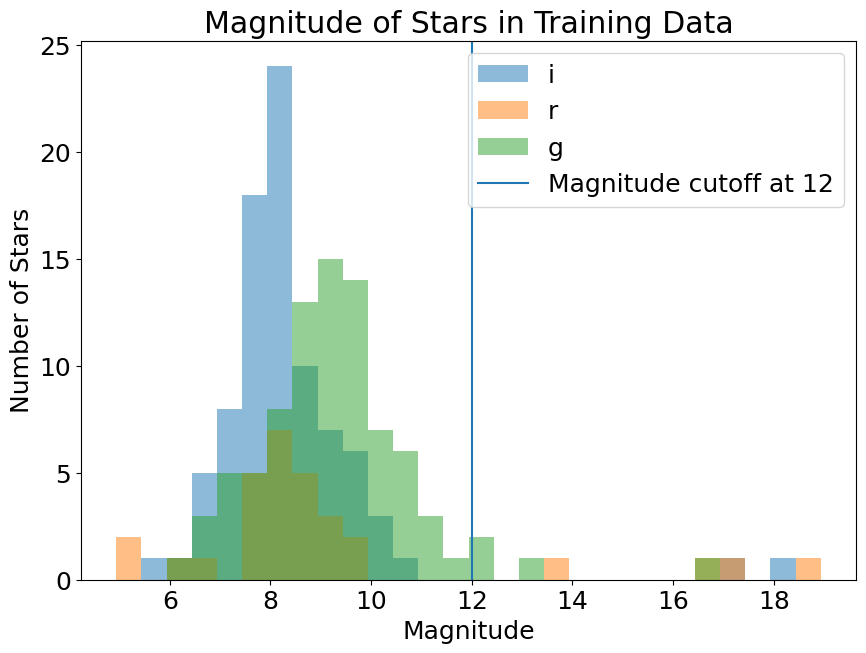}{0.45\textwidth}{}
       \fig{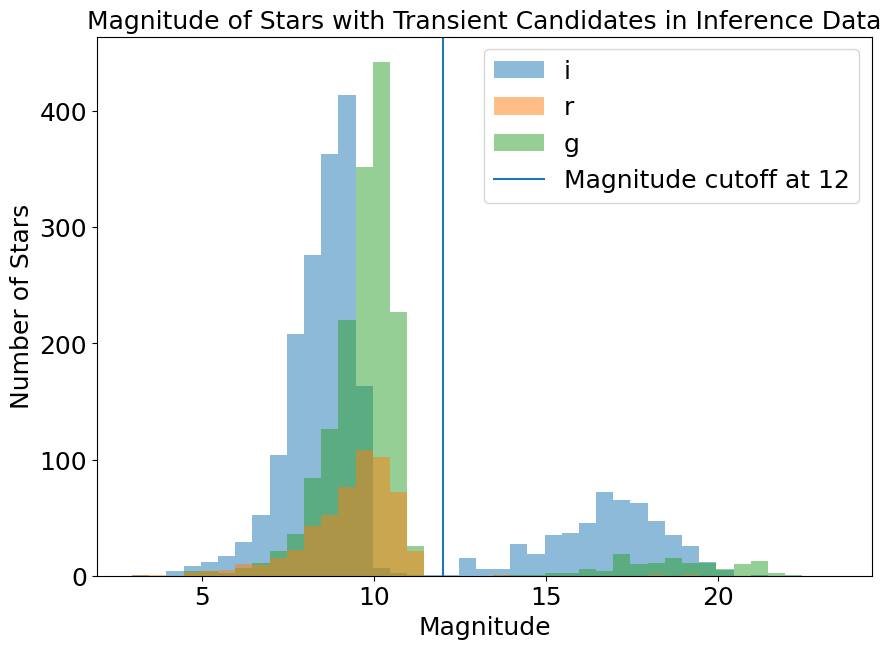}{0.45\textwidth}{}}
  \caption{Histograms showing the magnitudes of Gaia objects matching the streak initial positions, for both inference data and training data. Everything above magnitude 12 is assumed to be a false identification. The magnitudes are determined from the Gaia band that most closely matches the ZTF band that was observed: Gaia RP for ZTF i and ZTF r, and Gaia BP for ZTF g. }
  \label{fig:maglimit}
\end{figure}

Upon inspection of the results, any match where the Gaia source had a magnitude of $>12$ was determined to be a false identification, as is apparent from the clustering in \autoref{fig:maglimit}. Note that all sources that were observed in g-band with $G_{BP}< 12$ also have $G_{RP}< 12$, so $G_{RP}< 12$ can be used as an overall magnitude limit for our analysis.

Sources were categorized as stars, quasars, or white dwarfs according to their Gaia identifiers.

\section{Phase Space Plot Details}\label{sec:phasespace}

For references on each of the objects in \autoref{fig:phasespace}, see \autoref{tab:phasespace}.

\begin{table}[h!]
\footnotesize
\centering
\begin{tabular}{c|c}

\textbf{Phenomenon} & \textbf{References } 
 \\ 

\hline
Blazars & \citep{otero2024optical, goyal2021optical} \\ \hline 
Cataclysmic Variable Star Flickering & \citep{zamanov2016flickering} \\ \hline 
Cepheid Variable Stars & \citep{klagyivik2009observational} \\ \hline 
Kuiper Belt Occultations & \citep{nihei2007detectability} \\ \hline 
Oort Cloud Occultations & \citep{nihei2007detectability} \\ \hline 
LFBOT Flares & \citep{ho2023minutes} \\ \hline 

Active Galactic Nuclei & \citep{aranzana2018short, otero2024optical} \\ \hline 

Infant Supernovae and Shock-Breakout & \citep{bersten2018surge, modjaz2019new} \\ \hline 

Young Stellar Objects & \citep{stauffer2014csi} \\ \hline 

Micronovae & \citep{ilkiewicz2024classifying, scaringi2022triggering} \\ \hline 

Gamma Ray Burst Afterglows & \citep{panaitescu2011optical} \\ \hline 

Stellar Flares & \citep{yan2021characteristic, kowalski2024stellar} \\ \hline 

Magnetar Flares/Soft Gamma-Ray Repeaters & \citep{stefanescu2008very, dhillon2011first} \\ \hline 

Kilonovae & \citep{ascenzi2019luminosity} \\ \hline 

White Dwarf Modulations & \citep{burdge20208, marsh2016radio} \\ \hline 

X-Ray Binaries & \citep{zurita2003evidence, merloni2000magnetic, sako2018detection} \\ \hline

\end{tabular}
\caption{List of references for the phenomena in \autoref{fig:phasespace}.}
\label{tab:phasespace}
\end{table}

\vspace{.3 cm}

\bibliographystyle{aasjournal}


\bibliography{references}

\end{document}